\documentclass[usenatbib,useAMS]{mnras}
\usepackage{tablefootnote}
\usepackage{multirow}
\usepackage{xcolor}
\usepackage{hyperref}
\usepackage{graphicx}	
\usepackage{amsmath}	
\usepackage{amssymb}	
\usepackage{multicol}        
\usepackage{bm}		
\usepackage[normalem]{ulem} 
\usepackage{caption}      
\usepackage{url}
\usepackage{longtable}
\usepackage{pdfpages}

\title[Study of high-redshift quasars]{High-redshift quasars at $z \geq 3$ -- I. Radio spectra}

\author[Yu. Sotnikova et al.]{
Yu.~Sotnikova,$^{1}$\thanks{E-mail: lacertae999@gmail.com} 
A.~Mikhailov,$^{1}$
T.~Mufakharov,$^{1,2,3}$
M.~Mingaliev,$^{1,2}$
N.~Bursov,$^{1}$
T.~Semenova,$^{1}$
\newauthor
V.~Stolyarov,$^{1,2,4}$
R.~Udovitskiy,$^{1}$
A.~Kudryashova$^{1}$
and
A.~Erkenov$^{1}$
\\
$^{1}$Special Astrophysical Observatory of RAS, Nizhny Arkhyz, 369167, Russia\\
$^{2}$Kazan Federal University, 18 Kremlyovskaya St, Kazan 420008, Russia\\
$^{3}$Shanghai Astronomical Observatory, Chinese Academy of Sciences, Shanghai 200030, China\\
$^{4}$Astrophysics Group, Cavendish Laboratory, University of Cambridge,
      J J Thomson Avenue, Cambridge CB3 0HE, UK}

\date{Accepted 2021 July 16. Received 2021 July 09; in original form 2021 April 22}
\pubyear{2021}
\hypersetup{draft}
\begin{document}
\label{firstpage}
\pagerange{\pageref{firstpage}--\pageref{lastpage}}
\maketitle 

\begin{abstract}
We present the radio properties of optically selected quasars with $z\geq3$. The complete sample consists of 102 quasars with a flux density level $S_{1.4}\geq100$ mJy in a declination range $-35\degr \leq {\rm Dec} \leq +49\degr$. The observations were obtained in 2017--2020 using the radio telescope RATAN-600. We measured flux densities at six frequencies 1.2, 2.3, 4.7, 8.2, 11.2, and 22 GHz quasi-simultaneously with uncertainties of 9--31 per cent. The detection rate is 100, 89, and 46 per cent at 4.7, 11.2, and 22 GHz, respectively.
We have analysed the averaged radio spectra of the quasars based on the RATAN and literature data. We classify 46 per cent of radio spectra as peaked-spectrum, 24 per cent as flat, and none as ultra-steep spectra ($\alpha\leq-1.1$). The multifrequency data reveal that a peaked spectral shape (PS) is a common feature for bright high-redshift quasars. This indicates the dominance of bright compact core emission and the insignificant contribution of extended optically thin kpc-scale components in observed radio spectra. Using these new radio data, the radio loudness $\log~R$ was estimated for 71 objects with a median value of 3.5, showing that the majority of the quasars are highly radio-loud with $\log~R>2.5$. We have not found any significant correlation between $z$ and $\alpha$. Several new megahertz-peaked spectrum (MPS) and gigahertz-peaked spectrum (GPS) candidates are suggested. Further studies of their variability and additional low-frequency observations are needed to classify them precisely.
\end{abstract}


\maketitle

\section{Introduction}

High-redshift ($z\geq 3$) quasars are important to study because they provide information about the growth of supermassive black holes and the evolution of active galactic nuclei (AGNs) in the early Universe. Most of high-redshift and radio-loud objects are expected to be blazars, a subclass of AGNs with a relativistic jet pointing toward the observer \citep{1995PASP..107..803U}. Relativistically beamed sources dominate the high-redshift radio source population due to the Doppler boosting effect. The number of radio-loud ($R_{1.4}\geq 100$) quasars decreases with increasing redshift. For example, at $z>6$ only several radio-loud quasars are found, and radio emission is not detected at $z>7$ in optically detected quasars \citep{2007AJ....134..617W,2014AJ....147....6M}.

An observed radio spectrum of a blazar is a combination of a flat spectrum by a compact core and a steep spectrum by a kiloparsec jet. The relationship between the core and the jet power determines the shape of the observed radio spectrum based on their contribution. Hence, the majority of radio sources at high redshifts should have a flat spectrum. However, we observe the well-known steep-spectrum \citep{2008AJ....136..344M,2008A&A...484L..39F,2010A&A...524A..83F} and the gigahertz-peaked spectrum (GPS) population high-redshift radio-loud quasars \citep{1990MNRAS.245P..20O,2010A&A...524A..83F}. 

Currently, only a few high-redshift blazars are found at $z\geq 5$, for example SDSS J114657.79+403708.6 \citep{2010A&A...524A..83F,2014MNRAS.440L.111G}, SDSS J102623.61+254259.5 \citep{2012MNRAS.426L..91S,2015MNRAS.446.2921F}, SDSS J164854+460328 \citep{2019MNRAS.484..204C}, Q0906+6930 \citep{2004ApJ...610L...9R}, and PSO J047.4478+27.2992 \citep{2020A&A...635L...7B}. Their radio loudness is much less than that of quasars at redshifts $3 \leq z \leq 5$.

The most distant known blazar at $z=6.1$, PSO J047.4478+27.2992, was discovered by \cite{2020A&A...635L...7B} in 2020 and is considered a compact steep-spectrum or a megahertz-peaked spectrum (CSS/MPS) candidate based on the RATAN-600 and Very Large Array (VLA) measurements \citep{2020A&A...643L..12S,2021MNRAS.503.4662M}. It is a radio-loud blazar with a $\log R\sim 2.5$, which is higher by several orders of magnitude than for quasars at $z= 3$--5.

Several studies have compiled and studied statistically high-redshift blazars based on the optical (Sloan Digital Sky Survey (SDSS), e.g. \citealt{2013MNRAS.433.2182S}) or radio (Cosmic Lens All Sky Survey (CLASS)) observations (e.g. \citealt{2019MNRAS.484..204C}). A systematic study of 30 high-redshift ($z>4.5$) quasars was presented by \cite{2016MNRAS.463.3260C,2017MNRAS.467.2039C}, where multifrequency radio spectra were used to classify the spectral type of the sources and no preferable spectral type for high-redshift objects was found. \cite{1979A&A....80...13B}, \cite{1979A&AS...35..153T}, \cite{1980MNRAS.190..903L}, and \cite{2000A&AS..143..303D} demonstrated that sources with higher redshifts had steeper radio spectra. In some studies an ultra-steep spectrum is used either as an indicator of high redshift (e.g. \citealt{2001MNRAS.326.1563J, 2014A&A...569A..52S}) or as a reasonable tracer of intermediate-redshift ($z\geq 1$) galaxies \citep{2014MNRAS.443.2590S}.
There are only four radio-loud quasars known at $z > 6$ to date, PSO J172.3556+18.7734 at $z= 6.82$ is the most distant among them \citep{2021ApJ...909...80B}. It has a steep radio spectrum ($\alpha_{1.5-3}=-1.31$) and radio loudness about $R_{4400}\sim 70$.

Unfortunately, for most quasars at $z > 3$ observed data are available at very few radio frequencies. 
In this study, new RATAN-600 observed data for 102 high-redshift quasars are presented. In 2017--2020 we obtained radio spectra measured quasi-simultaneously at 1.2, 2.3, 4.7, 8.2, 11.2, and 22 GHz and compiled a catalogue of six-frequency flux densities. Combining our new observations with those from the literature, we investigated the radio spectra of quasars at $z\geq 3$. We estimated the radio luminosity and radio loudness for these objects and made an analysis for different redshift ranges. Spectral classification was performed and new GPS/MPS candidates were found.

\section{The sample}

\begin{figure}
\centerline{\includegraphics[width=\columnwidth]{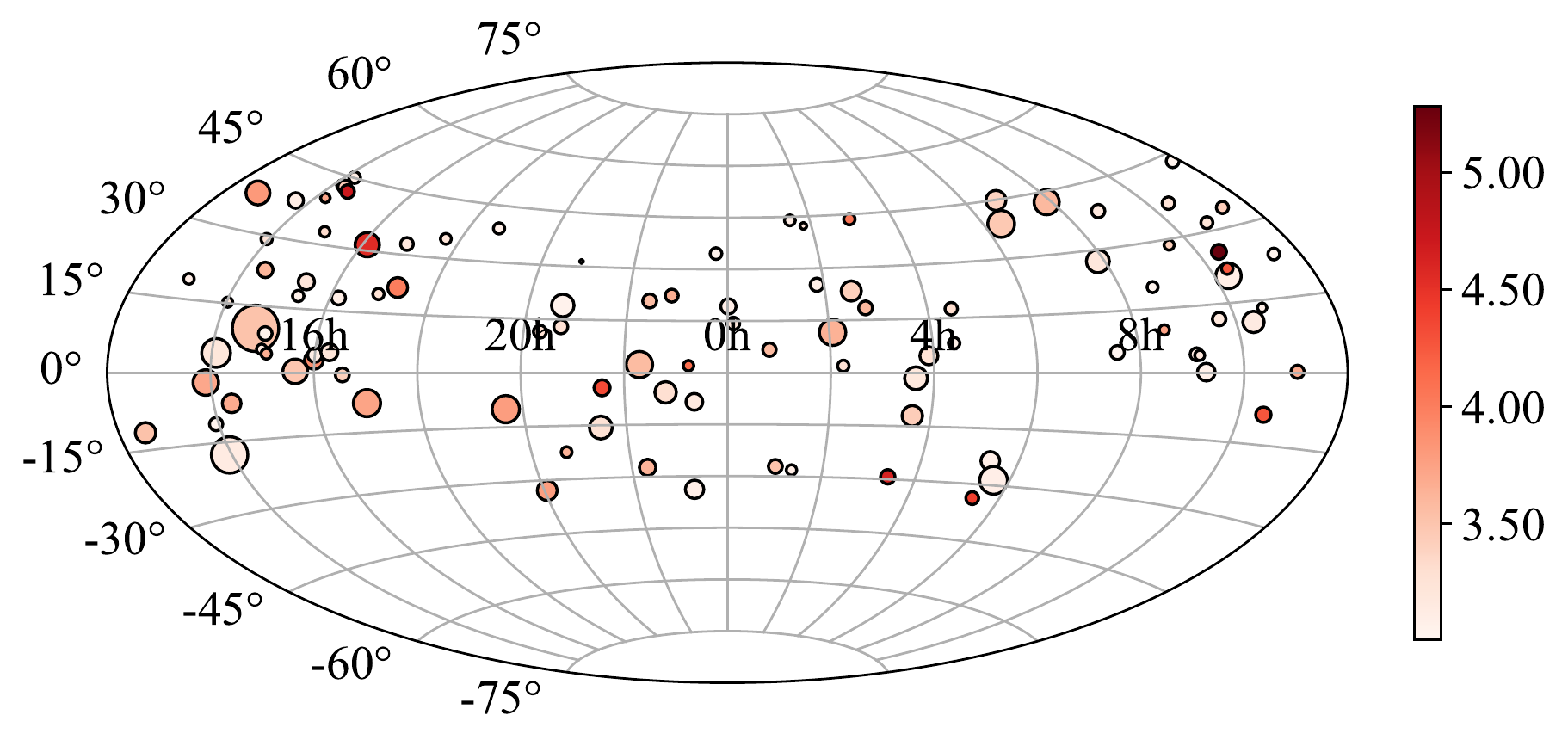}}
\caption{Hammer-Aitoff projection in galactic coordinates of the sky distribution of the quasars. The size of the circles corresponds to the flux density level at 1.4 GHz, and the color indicates the redshift value.}
\label{fig1}
\end{figure}

The complete sample consists of 102 quasars at $z\geq 3$ with a flux density level $S_{1.4}\geq100$ mJy in the declination range $-35\degr \leq {\rm Dec} \leq 49\degr$ (Fig.~\ref{fig1}).

\begin{figure}
\centerline{\includegraphics[width=\columnwidth]{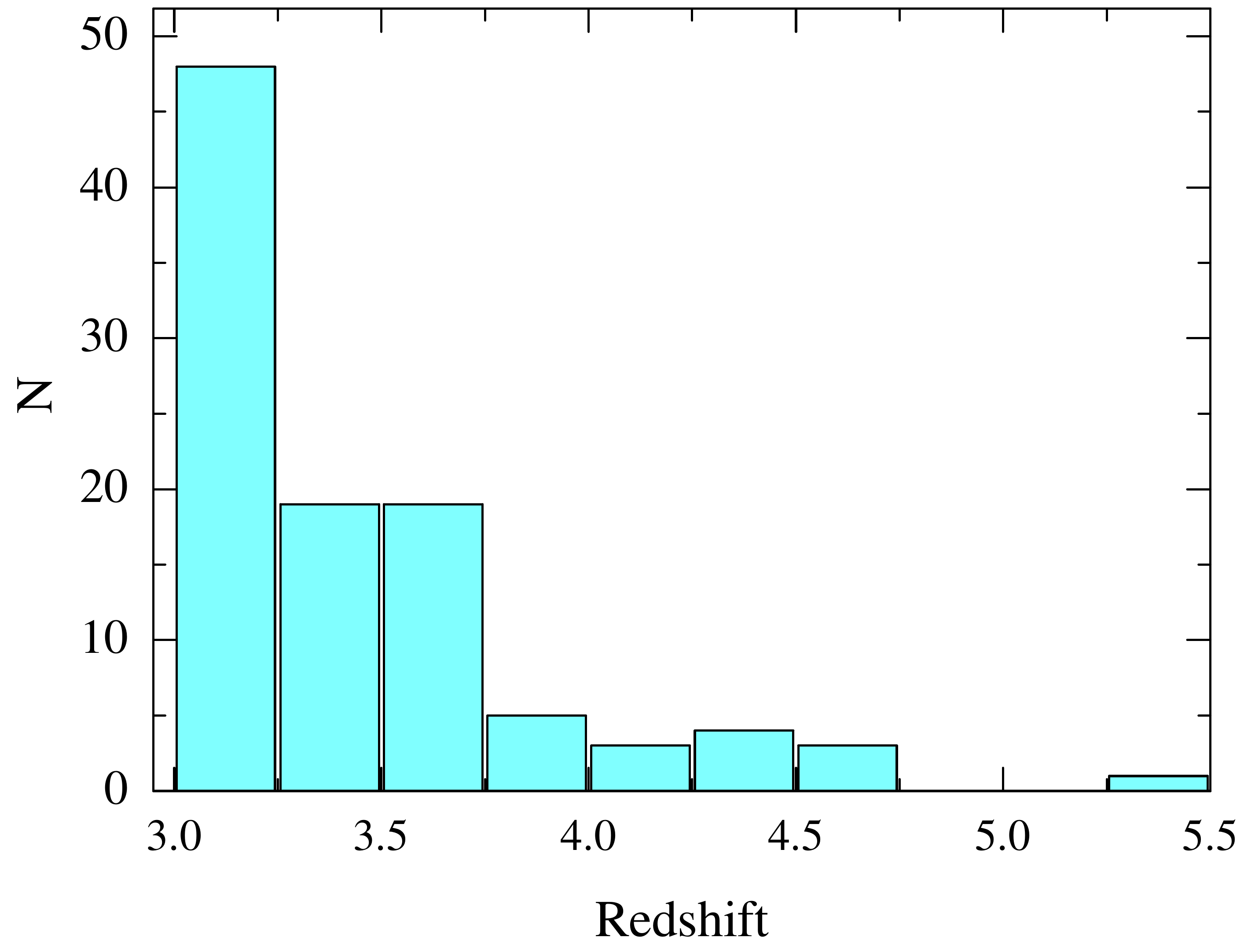}}
\caption{Redshift distribution for the sample.}
\label{fig2}
\end{figure}

\begin{figure}
\centerline{\includegraphics[width=\columnwidth]{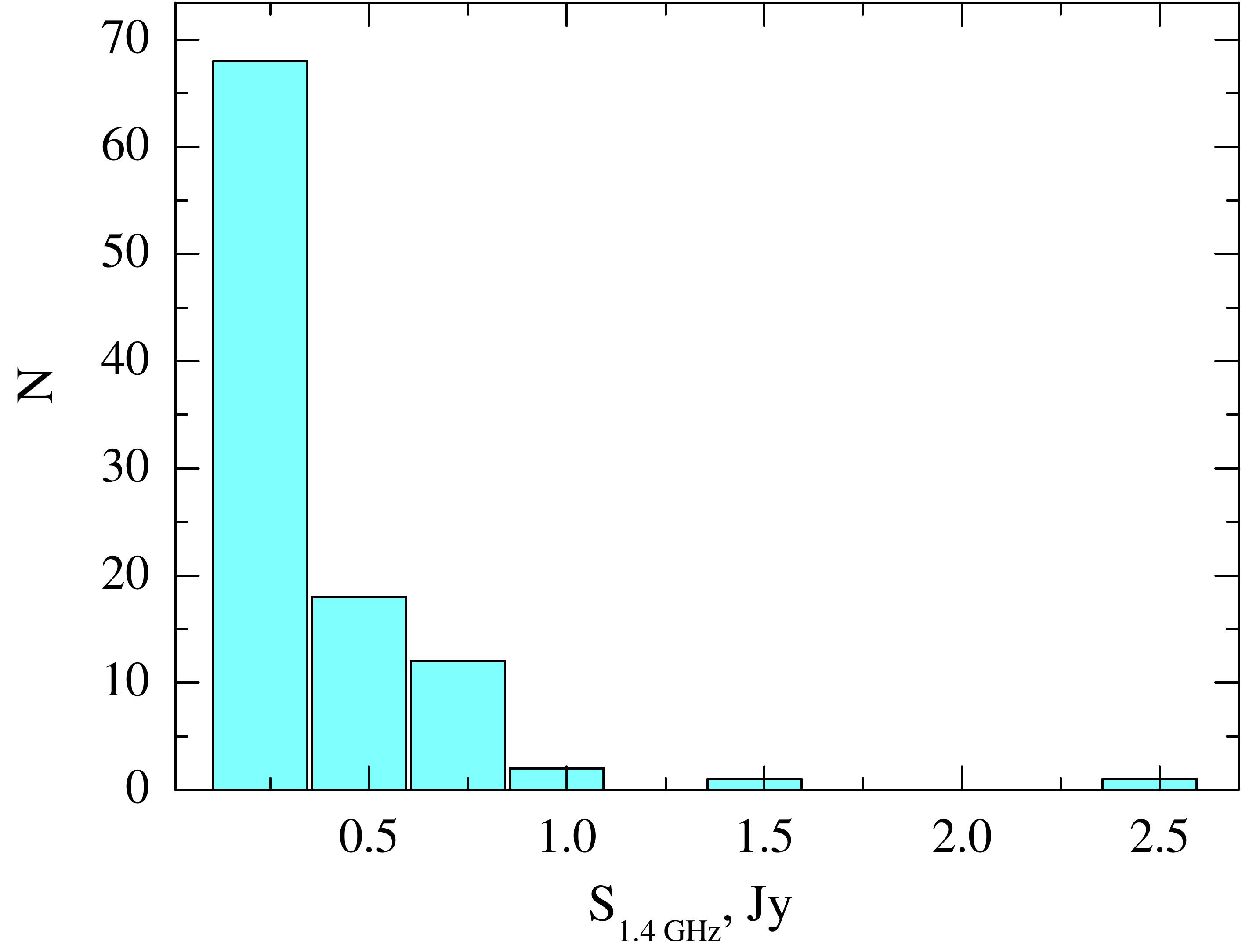}}
\caption{Flux density distribution for the sample at 1.4 GHz.}
\label{fig3}
\end{figure}

The spectroscopic redshifts were taken from the National Aeronautics and Space Administration (NASA)/Infrared Processing and Analysis Center (IPAC) Extragalactic Database (NED),\footnote{https://ned.ipac.caltech.edu} and the redshift median value was found to be 3.26. There are 10 quasars at $z\geq 4$ in the sample and only one, B2 1023+25, at $z> 5$ ($z= 5.28$; \cite{2012MNRAS.426L..91S}). The number of objects decreases exponentially with increasing redshift (Fig.~\ref{fig2}). There are 22 bright radio quasars with flux densities 0.5--2.4 Jy in the sample, and the median value of the flux density at 1.4 GHz is 0.22 Jy (Fig.~\ref{fig3}).

Table~\ref{tab:param} presents the list of the sources with their characteristics, where Column~1~is the source name, Column~2 is the redshift $z$, Column~3 is the average flux density at 4.7 GHz (RATAN-600 measurements), Column~4 is the radio luminosity at 4.7 GHz, Column~5 is the radio loudness $\log~R$, Columns~6 and 7 are the spectral indices, Column~8 is the radio spectrum type, and Column~9 is the blazar type.

The list includes 48 quasars that are classified as blazars according to the 5th Edition of the Roma-BZCAT catalogue \citep{2015Ap&SS.357...75M}. Almost all blazars (46) are flat-spectrum radio quasars (FSRQs), and two of them are blazars of an uncertain type.

For high-energy bands, we found that only seven sources have their matches in the \textit{Fermi}-LAT 4th catalog \citep{2020arXiv200511208B}, and 30 sources have X-ray photon fluxes available according to the NED database.

\section{Observations and data reduction}

We carried out the observations in transit mode with the RATAN-600 radio telescope \citep{1993IAPM...35....7P,2020gbar.conf...32S} at six frequencies between 1.2 and 22 GHz simultaneously. For a single transit observation, one can obtain an instantaneous spectrum of a source. RATAN-600 continuum radiometer parameters are presented in \citet{2019AstBu..74..348S}.

The sample was observed in 2017--2020. We observed the sources 3--15 times for each observing epoch to improve the signal-to-noise (S/N) ratio. The median number of the observing epochs is $N_{\rm obs}=9$ for the sample.

The observations were processed using the Flexible Astronomical Data Processing System (FADPS) standard data reduction software developed by \citet{1997ASPC..125...46V} for the broadband RATAN-600 continuum radiometers and the automated data reduction system \citep{2011AstBu..66..109T,2018AstBu..73..494T,2016AstBu..71..496U}. 
We used the following eight flux density secondary calibrators: 3C48, 3C138, 3C147, 3C161, 3C286, 3C295, 3C309.1, and NGC 7027.
The flux-density scale is based on measurements by \cite{1977A&A....61...99B} and \cite{2013ApJS..204...19P,2017ApJS..230....7P}; these are in good agreement, and their differences are within the measurement errors. Additionally we used traditional RATAN-600 flux density calibrators: J0240-23, J1154-35, and J0521+16 \citep{2019AstBu..74..497S}.
The measurements of the calibrators were corrected for angular size and linear polarisation, according to the data from \citet{1994A&A...284..331O} and \citet{1980A&AS...39..379T}.

The measurement uncertainties and the calibration procedure are described in \cite{2001A&A...370...78M} and \cite{2021MNRAS.503.4662M}.

\section{Results}

\subsection{Flux densities}
\label{flux}

The total number of observations is more than 1400, and the detection rates are 46 per cent and 100 per cent at 22 and 4.7 GHz, respectively (Table~\ref{tab:detection}). The number of sources with RATAN data available at five to six frequencies is 75. The median value of the S/N ratio varies from 9 to 17 at 4.7--22 GHz. At 1.2 and 2.3 GHz, the median S/N values are 25 and 31 due to strong radio frequency interference (RFI). The strong RFI contamination at 11.2 GHz is also the reason for the non-detection of objects in the declination range $-10\degr \leq {\rm Dec.} \leq 0\degr$.

\begin{figure}
\centerline{\includegraphics[width=.48\textwidth]{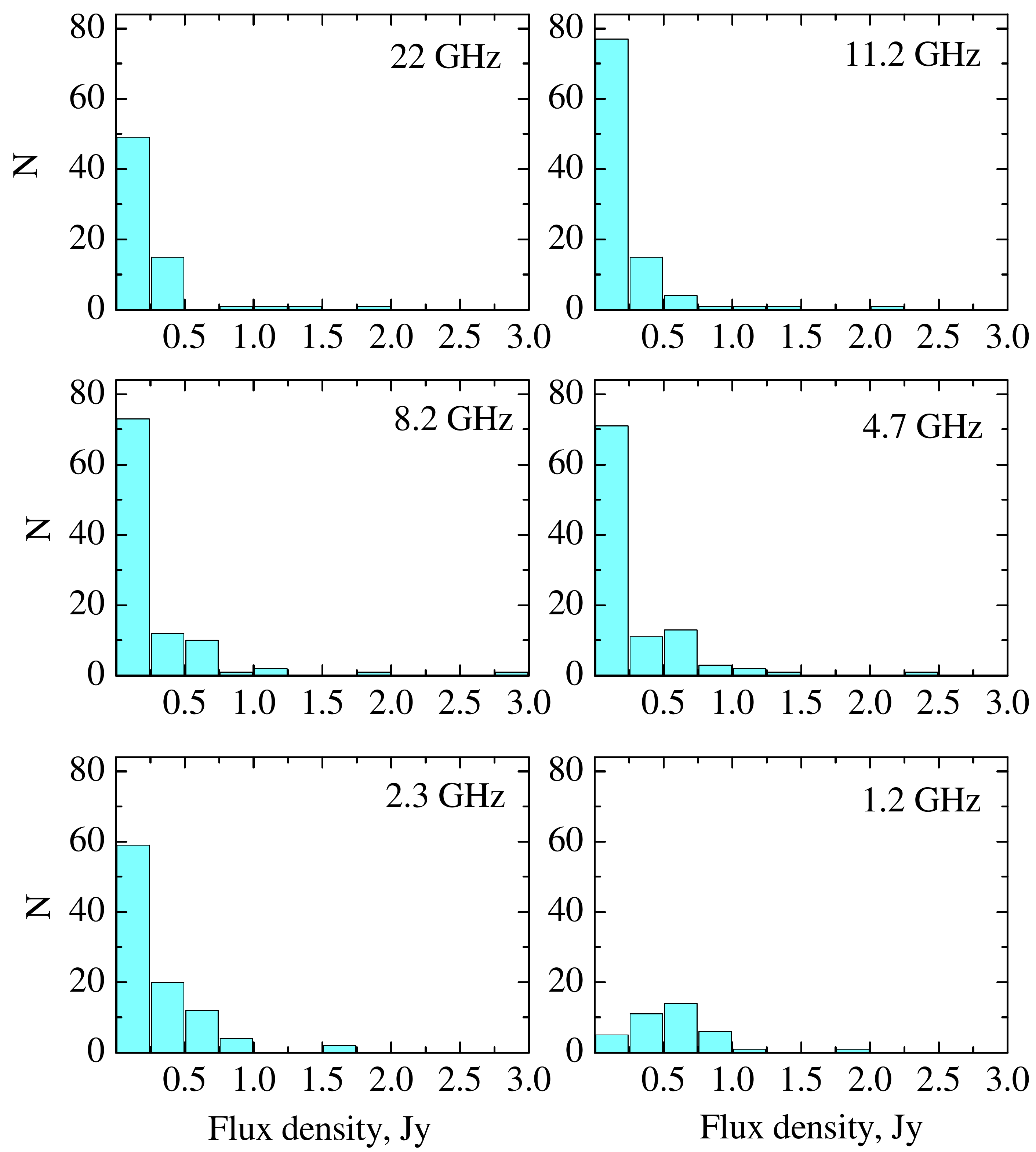}}
\caption{Averaged flux density distributions for the sample at RATAN frequencies.}
\label{fig4}
\end{figure}

The flux density standard errors are 5--20 per cent for 11.2, 8.2, and 4.7 GHz, and 10--35 per cent for 2.3, 1.2, and 22 GHz. Table~\ref{tab:detection} presents the detection rate statistics, median flux densities, and their uncertainties for different frequencies. Fig.~\ref{fig4} presents the average flux density distribution at the six frequencies.

\begin{table}
\caption{\label{tab:detection} Observational statistics: the mean value of the RATAN flux densities $\tilde{S}$ and their uncertainties $\tilde\sigma_{S}$, number of observations N$_{\rm obs}$, and detection rates. The standard deviations are given in parentheses.}
\centering
\begin{tabular}{rcrrr}
\hline
   Frequency  &   $\tilde{S}$ & $\tilde\sigma_{S}$ & N$_{\rm obs}$ & Detections  \\
          (GHz)    &     (Jy)          &  (\%)             &                   & (\%)  \\ 
\hline                                
22            & 0.56 (0.59) & 18 & 1418 & $46$  \\ 
11.2          & 0.41 (0.55) & 11 & 1418  & $89$  \\ 
8.2           & 0.32 (0.51) & 11 &  801  & $99$  \\ 
4.7           & 0.44 (0.52) & 9  &  1418 & $100$  \\ 
2.3           & 0.38 (0.37) & 31 &  801  & $82$   \\ 
1.2           & 0.77 (0.50) & 25 &  801  & $12$  \\ 
\hline
\end{tabular}
\end{table}

\begin{table*}
\caption{\label{tab:ratan}RATAN-600 flux densities for 102 quasars, measured simultaneously at six frequencies (1.2--22 GHz). A sample is shown here; the full version is available as online supporting material.}
\begin{tabular}{cccccccc}
\hline
NVSS name & JD & $S_{22}$ & $S_{11.2}$ & $S_{8.2}$ & $S_{4.7}$ & $S_{2.3}$ & $S_{1.2}$ \\
\hline
000108$+$191434 &  2457793 & --           & $0.168\pm0.016$ & $0.166\pm0.017$ &   $0.167\pm0.015$ &   $0.151\pm0.055$ &   --        \\
000108$+$191434 &  2457838 & $0.104\pm0.023$ & $0.134\pm0.014$ & $0.156\pm0.016$ &   $0.190\pm0.017$ &   $0.207\pm0.055$ &   --         \\
000108$+$191434 &  2457853 & $0.147\pm0.029$ & $0.129\pm0.013$ & $0.155\pm0.016$ &   $0.150\pm0.013$ &   $0.136\pm0.055$ &   --         \\
000108$+$191434 &  2458027 & $0.146\pm0.029$ & $0.135\pm0.014$ & $0.150\pm0.015$ &   $0.164\pm0.014$ &   $0.207\pm0.055$ &   $0.339\pm0.096$\\
000108$+$191434 &  2458104 & $0.111\pm0.024$ & $0.126\pm0.013$ & $0.141\pm0.015$ &   $0.170\pm0.015$ &   $0.167\pm0.055$ &  --        \\
000108$+$191434 &  2458192 & --          & $0.108\pm0.012$ & $0.133\pm0.014$ &   $0.157\pm0.014$ &   $0.144\pm0.055$ &   --         \\
000108$+$191434 &  2458236 & --          & $0.129\pm0.013$ & $0.151\pm0.016$ &   $0.153\pm0.013$ &   $0.167\pm0.055$ &   --         \\
000108$+$191434 &  2458309 & $0.112\pm0.024$ & $0.117\pm0.012$ & $0.152\pm0.016$ &   $0.168\pm0.015$ &   $0.194\pm0.055$ &   --  \\
000108$+$191434 &  2458530 & --          & $0.094\pm0.010$ & $0.107\pm0.012$ &   $0.134\pm0.012$ &   $0.153\pm0.055$ &   --         \\
000108$+$191434 &  2458717 & $0.103\pm0.022$ & $0.119\pm0.012$ & $0.118\pm0.013$ &   $0.139\pm0.012$ &   $0.125\pm0.056$ &   $0.181\pm0.067$ \\
000108$+$191434 &  2458771 & --          & $0.088\pm0.010$ & $0.106\pm0.012$ &   $0.120\pm0.011$ &   $0.122\pm0.056$ &   --   \\
\hline
\end{tabular}
\end{table*}

Table~\ref{tab:ratan} presents the flux densities, where Columns 1 and 2 are the source name and the averaged observing epoch (JD), Columns 3--14 are the flux densities and their uncertainties (Jy). The data is distributed in the VizieR Information System.\footnote{https://vizier.u-strasbg.fr/viz-bin/VizieR} The RATAN-600 flux densities and the instantaneous spectra for some quasars are published in the BL Lac database\footnote{www.sao.ru/blcat/} maintained by the Special Astrophysical Observatory. The database is constantly updated with new data and is freely available \citep{2014A&A...572A..59M}.

\subsection{Radio spectra and classification}

Figs.~\ref{fig:spectra}--\ref{fig:spectra4} show the broad-band radio spectra of the quasars. We compiled the spectra using both RATAN (red circles) and CATS (black circles) data \citep{2005BSAO...58..118V, 1997BaltA...6..275V}. The RATAN-600 measurements cover the time period 2017--2020, except for several objects which were observed from 2014. The CATS data cover a time period of several dozens years, and the main data are represented by the National Radio Astronomy Observatory (NRAO) VLA Sky Survey (NVSS: \citet{1998AJ....115.1693C}), Faint Images of the Radio Sky at Twenty cm (FIRST: \citet{1994ASPC...61..165B}), Westerbork Northern Sky Survey (WENSS: \citet{1997A&AS..124..259R}), Green Bank 6-cm survey (GB6: \citet{1996ApJS..103..427G}), Australia Telescope 20 GHz Survey (ATCA20: \citet{2010MNRAS.402.2403M}), and VLA measurements.
Low-frequency data are well presented by the GaLactic and Extragalactic All-sky Murchison Widefield Array (GLEAM) survey at 72--231 MHz (2013--2014; \cite{2017MNRAS.464.1146H}) and the Giant Metrewave Radio Telescope Sky Survey (TGSS) at 150 MHz (2015, \cite{2017A&A...598A..78I}); the data were obtained almost simultaneously with the RATAN-600 observations.

We analysed the averaged radio spectra of quasars compiled from the RATAN and CATS data. The spectral parameters were estimated by fitting with polynomials, which were calculated by weight averaging of measurements in the FADPS software package \citep{1997BaltA...6..275V}. The resulting approximations are shown by the green dotted lines in Figs.~\ref{fig:spectra}--\ref{fig:spectra4}. We defined the spectral index from the power-law $S_{\nu} \propto \nu^{\alpha}$ ($S_{\nu}$ is a flux density at the frequency $\nu$, and $\alpha$ is a spectral index).

Table~\ref{tab:type} presents the criteria based on which the spectral classification was made. We determined the spectral indices $\alpha_{\rm low}$ and $\alpha_{\rm high}$ for peaked and upturn spectra for the frequencies where the spectral slope changes its sign from positive to negative or vice versa. For the rest of the spectra, the low-frequency spectral index was calculated between the TGSS, GLEAM, WENSS, and other decimeter-wavelength measurements. The high-frequency spectral index was estimated between centimeter catalog data points. We considered a spectrum to be flat if the spectral index $-0.5 \leq \alpha \leq 0$, and inverted if $\alpha > 0$. The spectrum is assumed to be complex for indeterminate shapes with two or more maxima or minima. The spectrum is called steep or ultra-steep for $-1.1 < \alpha < -0.5$ and $\alpha \leq -1.1$, respectively.

\begin{figure}
\centerline{\includegraphics[width=\columnwidth]{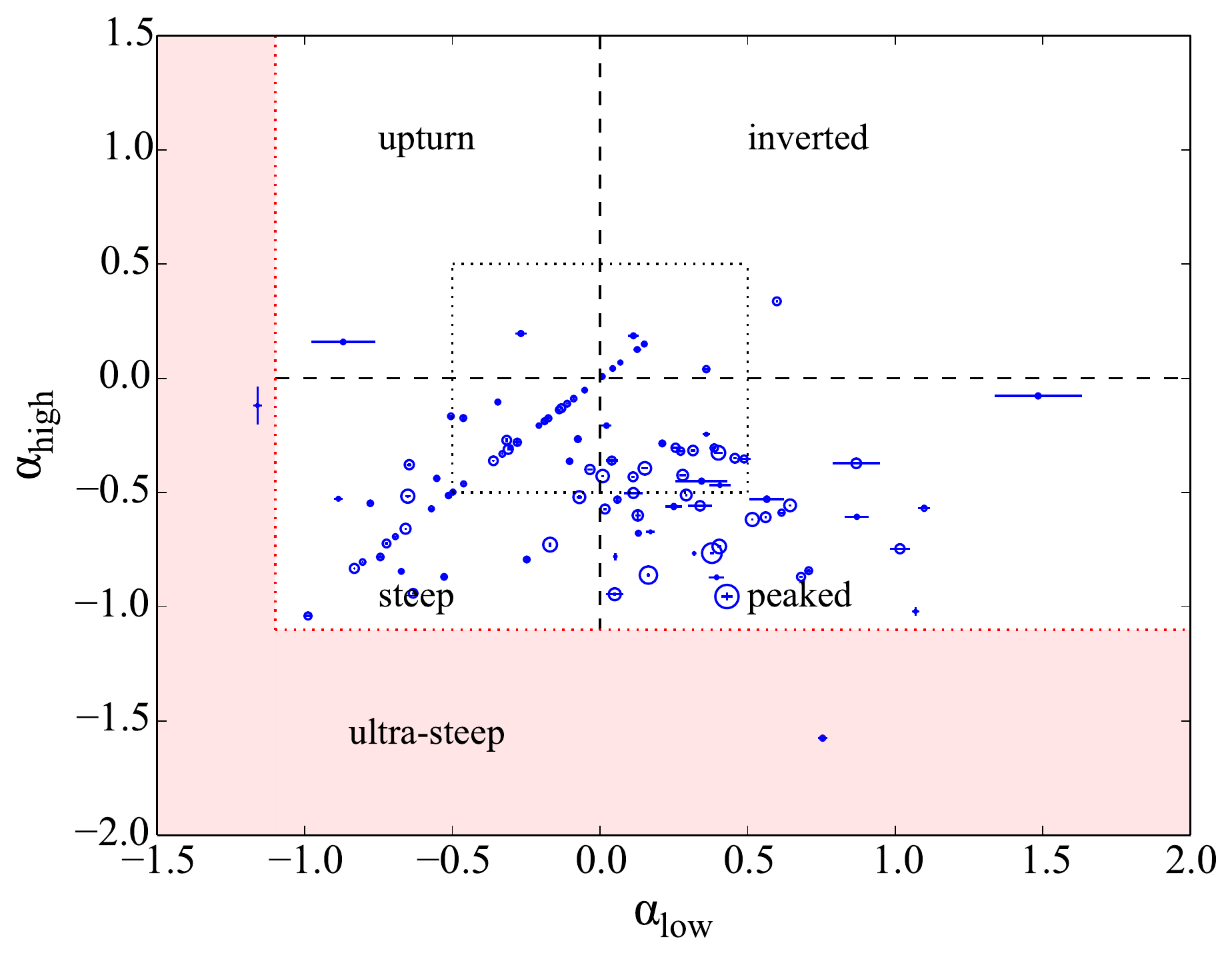}}
\caption{Two-color diagram. The dotted square indicates the 'blazar box', the dashed lines mark the zero spectral indices, and the ultra-steep spectrum sources area is presented with a pink color. The point size is proportional to the radio luminosity at 4.7 GHz.}
\label{fig5}
\end{figure}

\begin{figure}
\centerline{\includegraphics[width=.48\textwidth]{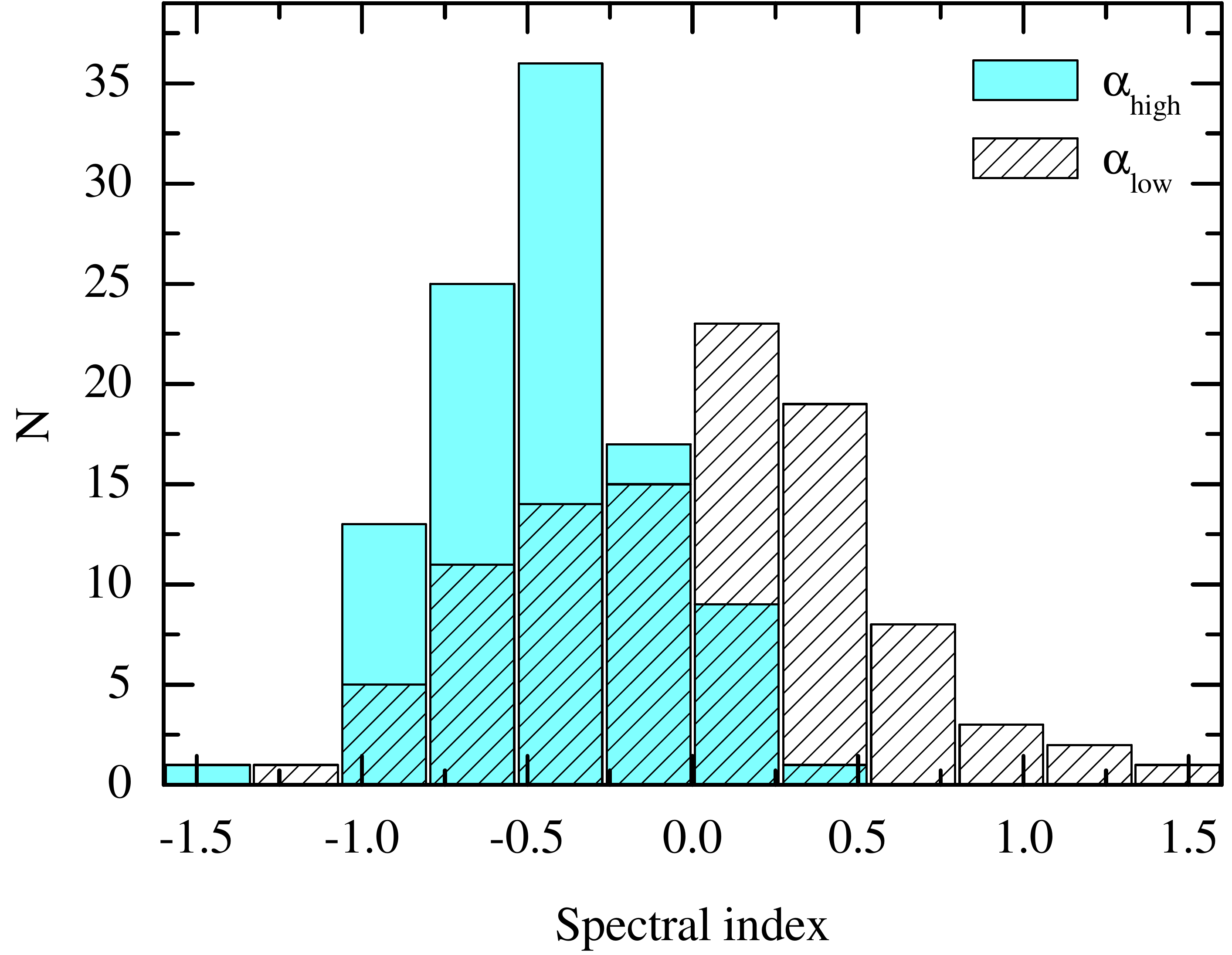}}
\caption{Distribution of the low-frequency (shaded) and high-frequency (colored) spectral indices for the averaged radio spectra.}
\label{fig6}
\end{figure}

Fig.~\ref{fig5} presents the spectral-type distribution in the convinient form of a two-color diagram in which $\alpha_{\rm low}$ is plotted against $\alpha_{\rm high}$. The point size is proportional to the radio luminosity at 4.7 GHz. The 'blazar box' corresponds to the spectral index range from $-$0.5 to 0.5. The median values of $\alpha_{\rm low}$ and $\alpha_{\rm high}$ are $-$0.01 (0.52) and $-$0.44 (0.33), respectively (Fig.~\ref{fig6}). Thus, the majority of the radio spectra steepens at higher frequencies.

We found 47 (46 per cent) of quasars had peaked radio spectra (PS) with a turnover in their average spectrum (Table~\ref{tab:PS}). There are 17 sources with classical \citep{1997A&A...321..105D} GPS spectral features and 13 quasars with a peak at MHz frequencies (MPS). We present 24 new GPS and MPS candidates in Table~\ref{tab:PS}.

A large part of the sample (24 per cent) has flat spectra and 15 per cent has steep ones. A small number of the sources have complex, upturn, and inverted spectra. 

We have not found ultra-steep radio spectra. The spectrum of PKS 2313-330 has an ultra-steep part with $\alpha_{4.7{\text{--}}11.2}=-1.58\pm0.09$ and is classified as PS. The spectral index was determined based on the multi-epoch RATAN observations in 2015--2020 at four frequencies simultaneously. For the second quasar, 87GB 012817.1+434250, we found an ultra-steep spectral part at low frequencies with $\alpha_{1.4{\text{--}}4.7}=-1.16\pm0.02$. This object is classified as having a complex spectrum. Its ultra-steep spectral part was obtained from two non-simultaneous measurements (NVSS and RATAN) on a time-scale of 20 years.

For nine quasars, the radio spectra are well described by a sum of two spectral components at low (LFC) and high (HFC) frequencies \citep{2002PASA...19...83K,2020AdSpR..65..745K}, which are denoted by '*' in Table~\ref{tab:param}. LFC indicates an optically thin synchrotron emission of extended kpc-scale structures with a peak frequency much less than 1 GHz due to its characteristic size \citep{1963Natur.199..682S}. HFC dominates at frequencies of several GHz and more and is associated with the pc-scale jet.

\begin{table}
\begin{small}
\caption{\label{tab:type}Spectral types in the sample.}
\centering
\begin{tabular}{lccc}
\hline
 Type & Criteria  &  N & {\%} \\
\hline
Peaked  & $\alpha_{\rm low}>0$, $\alpha_{\rm high}<0$ & 47 & 46  \\ 
Flat    & $-0.5 \leq \alpha \leq 0$  &  25  & 24 \\ 
Inverted  & $\alpha>0$ &  8  & 8 \\ 
Upturn  & $\alpha_{\rm low}<0$, $\alpha_{\rm high}>0$ &  2  & 2 \\ 
Steep  & $-1.1 < \alpha < -0.5$ & 15 & 15 \\ 
Ultra-steep  & $\alpha \leq -1.1$ & 0  & 0 \\ 
Complex  & Two or more maxima/minima & 5  &  5 \\ 
\hline
\end{tabular}
\end{small}
\end{table}

\subsection{The average spectrum}
To study the distant radio sources population, we used continuum radio spectra averaged over the objects located within close redshift intervals $\Delta z = 0.1$. The spectrum of each object was preliminarily recalculated into the rest frame using the $(1+z)$ correction. The spectra of individual objects were then approximated by a polynomial of the first or second degree. Finally, the average spectrum of all objects in each bin was calculated using the formula \citep{2018AstBu..73..393V}
\begin{equation}
\log\,S(\nu_{j}) = \frac{1}{N}\sum\limits_{i=1}^N \log\,f_{i}(\nu_{j}),
\end{equation}
where $\nu_{j}$ is the set of frequencies in the approximating data set for the $i$the source, $f_{i} (\nu_{j})$ is the flux density calculated for the $i$th source approximation at a frequency $\nu_{j}$, and $N$ is the number of sources. The averaging procedure was carried out using the $spcalc$ utility of the FADPS data processing system. As a result, we have calculated eight average spectra in the range $z = 3.0$--3.8. The mean error of the averaged spectrum in each redshift bin is determined by the 
quadratic sum of spectrum approximation uncertainties. The uncertainty of the spectrum approximation of an individual object was calculated using errors of polynomial fitting coefficients by a standard method. At higher redshift intervals, there are only a few objects, and the averaging procedure was not applied.

\begin{figure}
\centerline{\includegraphics[width=\columnwidth]{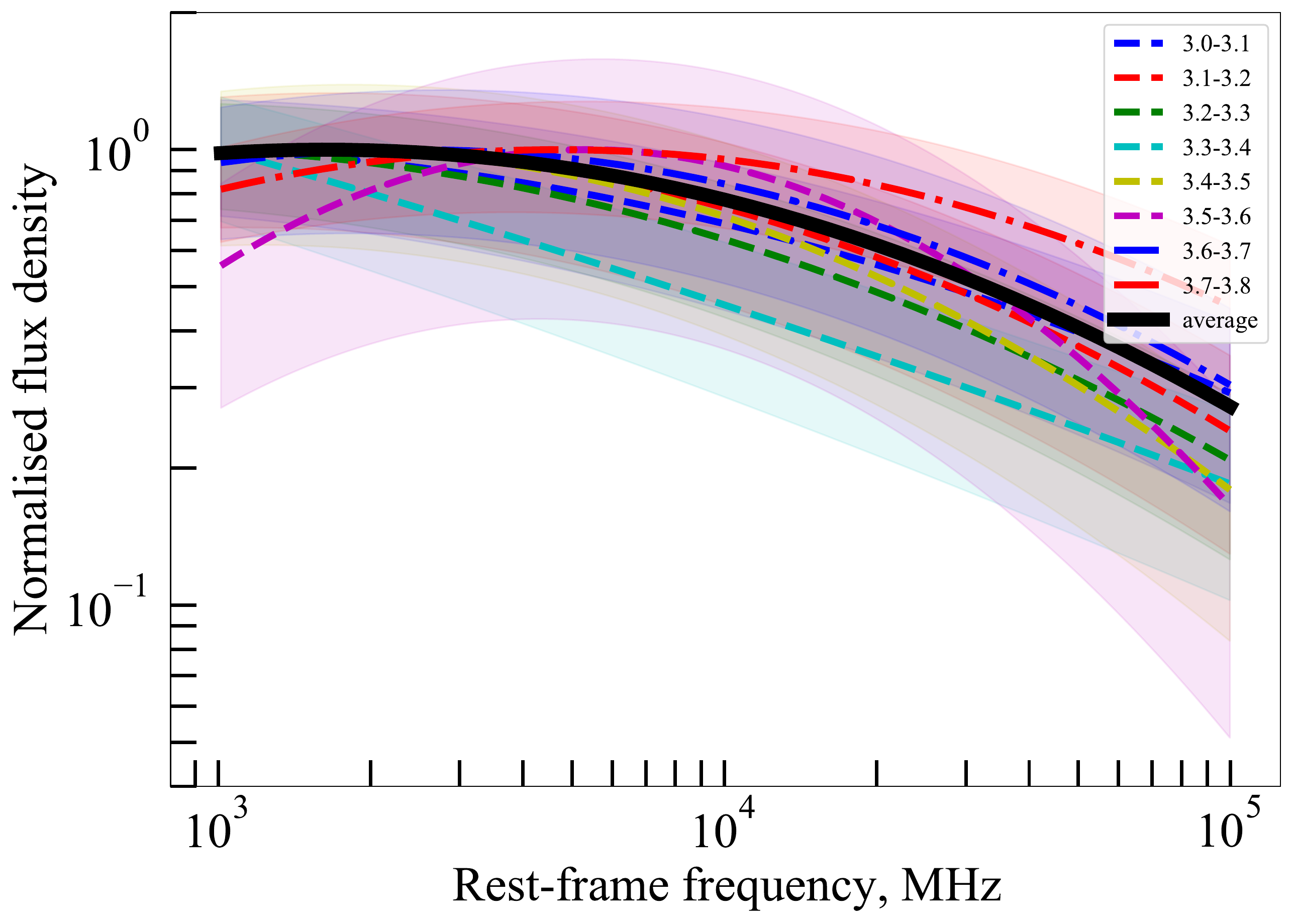}}
\caption{Average spectra of the sources normalised by the peak flux density. The coloured lines relate to the redshift bins $z = 3.0$--3.8 and corresponding strips represent 3$\sigma$ uncertainties. The black line is the average spectrum of all sources at $z=3.0-3.8$.}
\label{fig7}
\end{figure}

Fig.~\ref{fig7} presents the normalised average radio spectra for the redshift bins $\Delta z=0.1$. The cumulative radio spectrum is shown by the black line. A 99 per cent confidence intervals (3$\sigma$) are shown as strips with different colors.

The eight average spectra in the redshifts range 3.0--3.8 show a similar shape within their uncertainties. The values of low and high spectral indices for the cumulative radio spectrum are $\alpha_{\rm low}$=0.20$\pm$03 and $\alpha_{\rm high}$=-0.40$\pm$0.02, respectively. The peak frequency for the average spectrum is about 1.6 GHz in the rest frame.

\subsection{$z$--$\alpha$ correlation}
\begin{figure}
\centerline{\includegraphics[width=\columnwidth]{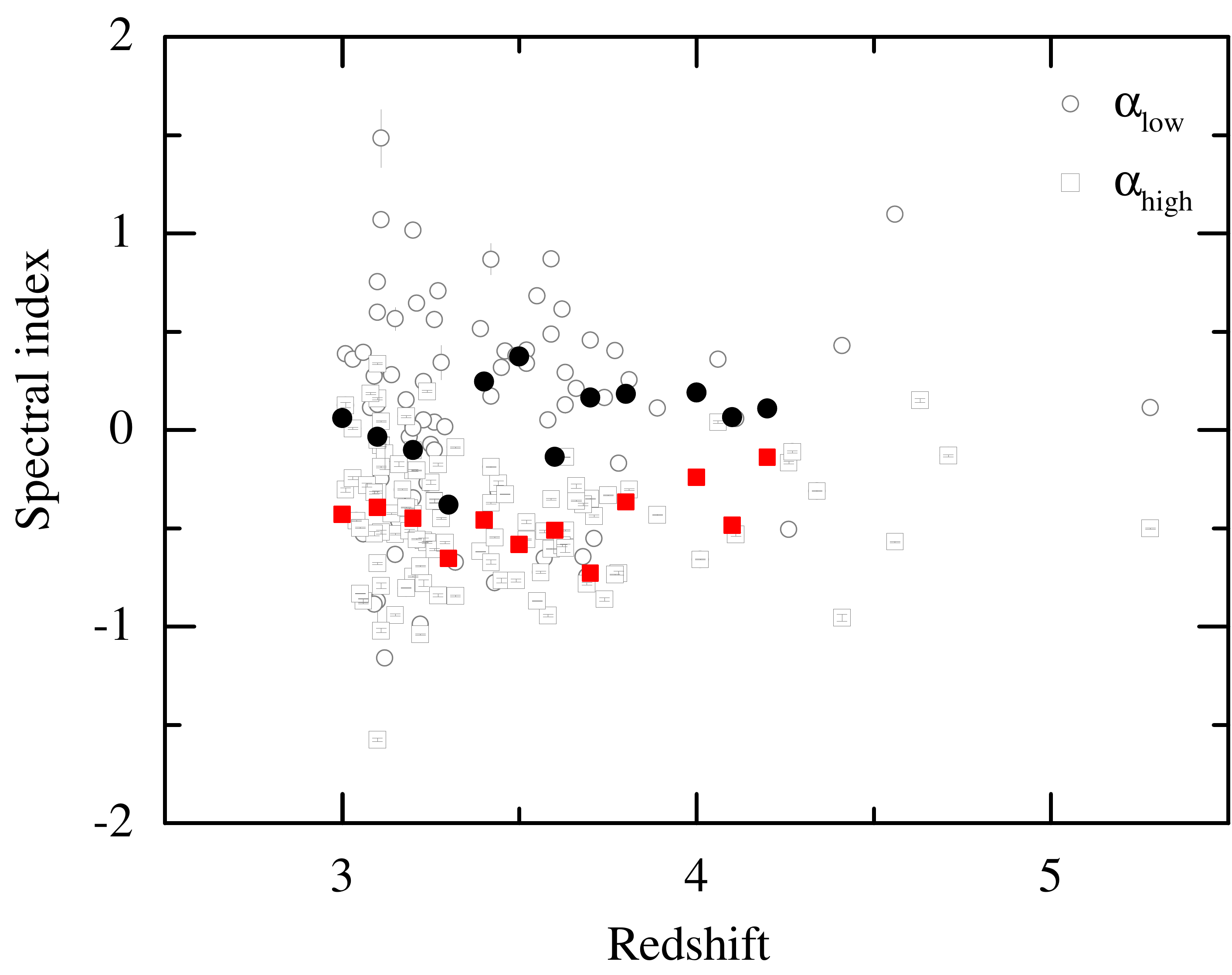}} 
\caption{$\alpha$ versus $z$ plot for low and high frequencies. The filled symbols represent the median values of the spectral indices for redshift bins of 0.1, from 3.0 to 4.2.}
\label{fig8}
\end{figure}

We have not found any significant correlation between redshifts and spectral indices at low and high frequencies (Fig.~\ref{fig8}). The Pearson correlation coefficient is $r=0.07$ for $\alpha_{\rm low}-z$ and $r=0.01$ for $\alpha_{\rm high}-z$, with 46 and 95 per cent of probability that these weak correlations appear by chance and not weighty, although this dependence was reported for radio sources in several studies (e.g. \citealt{1979A&A....80...13B,1980MNRAS.190..903L,2000A&AS..143..303D}).
The result does not contradict the existing theories, explaining $z$--$\alpha$ correlation by the interaction of the radio lobes with the dense intergalactic medium at $z\sim~2$--3. Our sample contains distant ($z>~3$) and very compact radio sources, mostly with peaked radio spectra, for which the core emission dominates at the GHz frequencies and the influence of radio lobes is negligible \citep{2012MNRAS.420.2644K}.

\subsection{Radio luminosity}
We used the $\Lambda$CDM cosmology with $H_0 = 67.74$ km s$^{-1}$ Mpc$^{-1}$, $\Omega_m$=0.3089, 
and $\Omega_\Lambda$=0.6911 \citep{2016A&A...594A..13P} to estimate the radio luminosity at 4.7 GHz:
 \begin{equation}
L_{\nu} = 4 \pi D_{\rm L}^2 \nu S_{\nu} (1+z)^{-\alpha -1}, 
\end{equation}
where $\nu$ is the frequency, $S_{\nu}$ the measured flux density, 
$z$ the redshift, $\alpha$ the spectral index calculated at 4.7 GHz by linear approximation of the spectrum between 1 and 8 GHz, and $D_{\rm L}$ the luminosity distance.

\begin{figure}
\centerline{\includegraphics[width=\columnwidth]{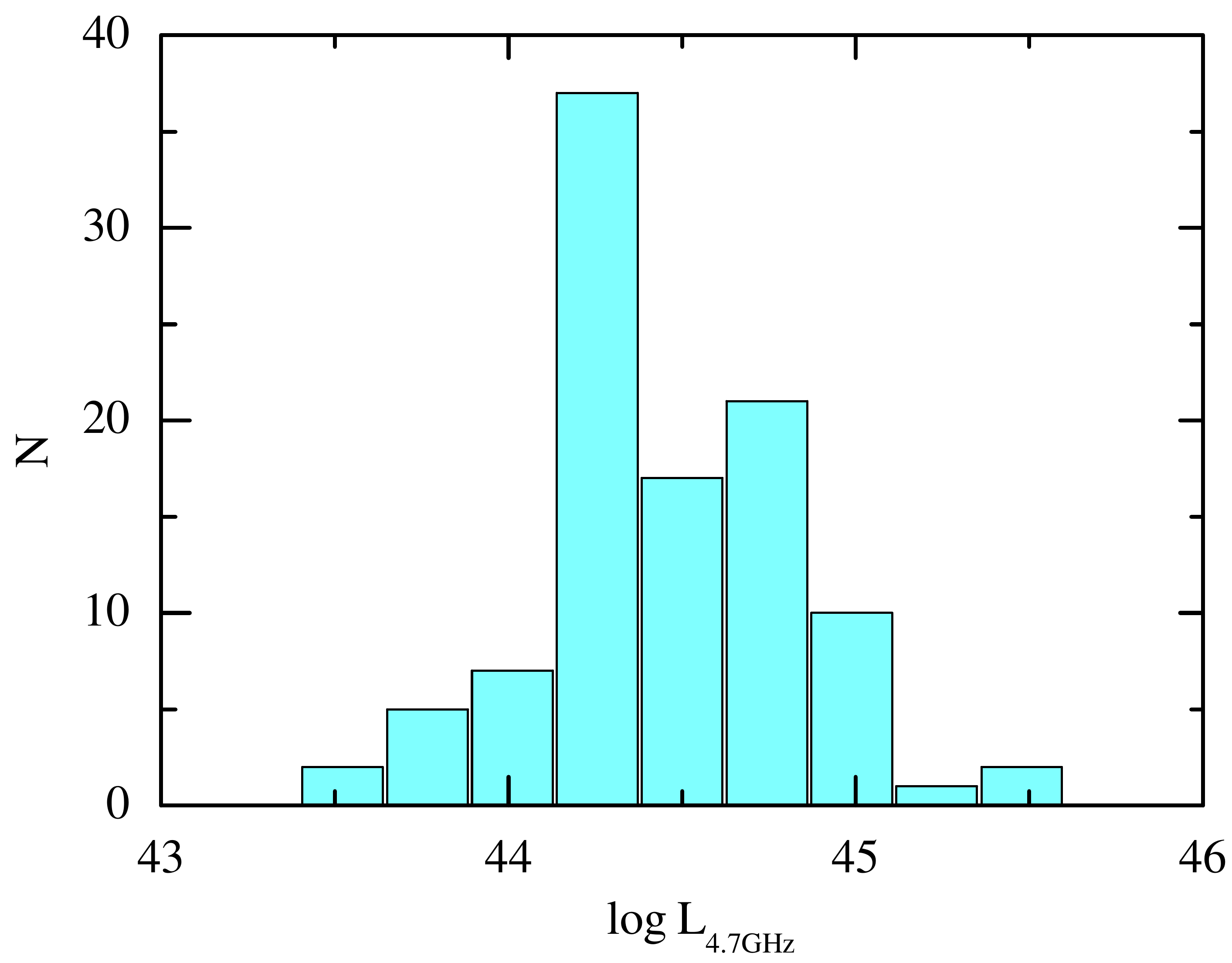}}
\caption{Radio luminosity distribution.}
\label{fig9}
\end{figure}

\begin{figure}
\centerline{\includegraphics[width=\columnwidth]{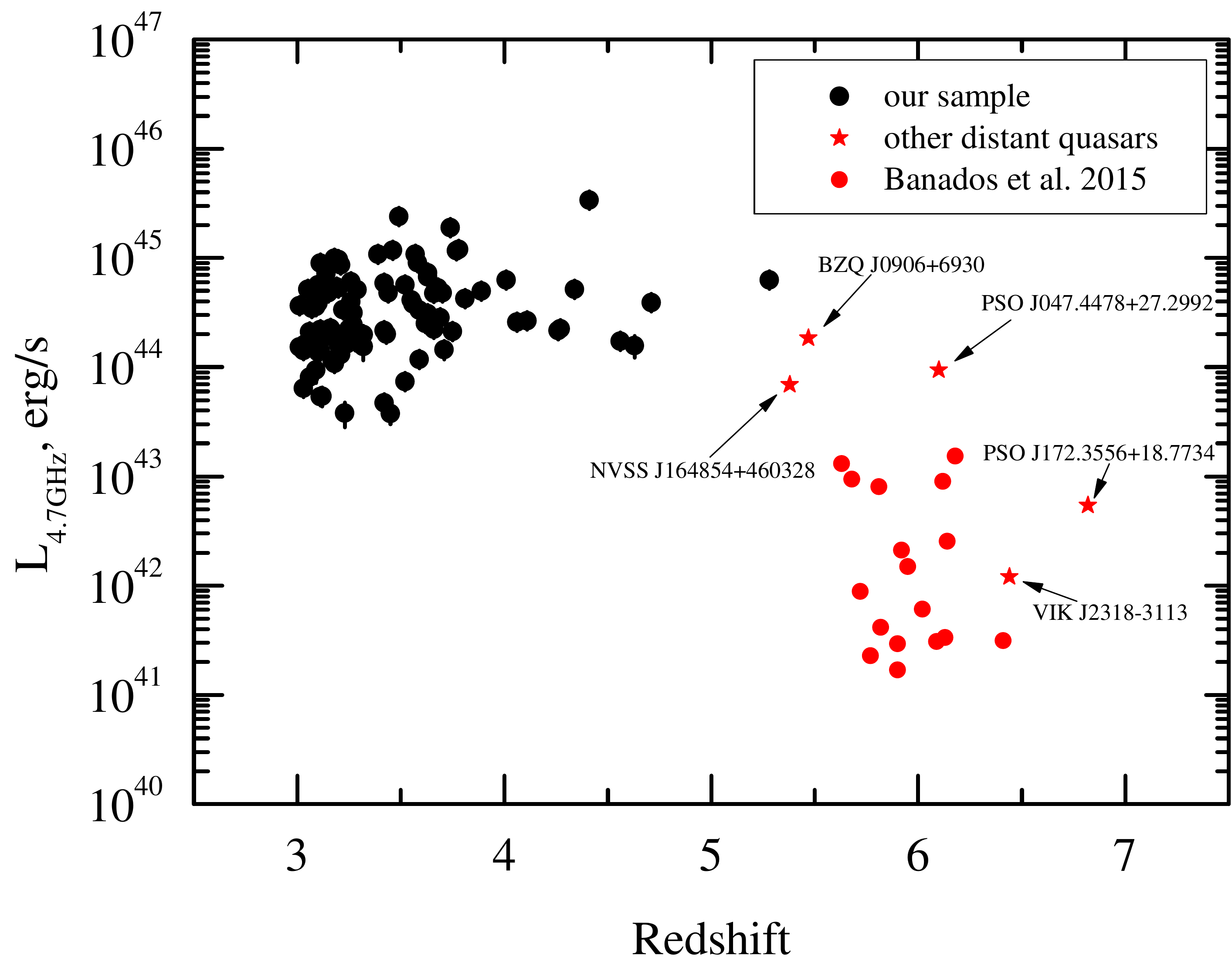}}
\caption{Redshift versus luminosity. The sample of distant blazars from \citet{2015ApJ...804..118B} is shown with the red circles and other five distant quasars from the literature are marked with the red stars.}
\label{fig10}
\end{figure}

The median value of the radio luminosity for our sample is $\sim 2 \times 10^{44}$ erg s$^{-1}$, and the source J0525$-$3343 at $z=4.41$ has the highest luminosity of $3.38 \times 10^{45}$ erg s$^{-1}$. Figs~\ref{fig9} and \ref{fig10} show the luminosity distribution and the redshift versus luminosity plot, respectively. 

In Fig.~\ref{fig10} we compare the luminosities of our sample with the luminosities calculated for the 17 distant quasars at $z>5.5$ selected from the FIRST and Pan-STARRS1 surveys cross-match in \citet{2015ApJ...804..118B}. Also, we draw in the figure five individual sources: BZQ J0906+6930 at $z=5.47$, PSO J047.4478+27.2992 at $z=6.1$ and NVSS J164854+460328 at $z=5.38$ \citep{2021MNRAS.503.4662M}, VIK J2318-3113 at $z=6.44$ \citep{2021arXiv210111371I}, and PSO J172.3556+18.7734 at $z=6.82$ \citep{2021ApJ...909...80B}. 
The sample of \citet{2015ApJ...804..118B} consists of the sources with the 1.4 GHz flux densities in the range 0.035--3.04 mJy, and accordingly their luminosity values are two orders of magnitude lower compared to the values in our sample.

\subsection{Radio loudness}
The radio loudness is defined as
\begin{equation}
R = \frac{S_{\nu, {\rm radio}}}{S_{\nu, {\rm opt}}}
\end{equation}
where $S_{\nu,{\rm radio}}$ is the estimated radio flux density at 4.7 GHz and $S_{\nu, {\rm opt}}$ is the optical flux density corresponding to the filter $B$ or $g$.

Optical data were taken from the SIMBAD database,\footnote{https://simbad.u-strasbg.fr/simbad/} where $B$ or $g$ magnitudes are presented for 71 of 102 objects. The magnitudes were transformed into flux densities using the Pogson law.
We adopted $S_{\nu,{\rm opt}} = 3631$ Jy for $g$ = 0,\footnote{\url{https://www.sdss.org/dr12/algorithms/fluxcal/\#SDSStoAB}}, $S_{\nu,{\rm opt}} = 4260$ Jy for $B = 0$ \citep{1979PASP...91..589B}, and the optical spectral index $\alpha=-0.3$ \citep{2019A&A...630A.110G}.

\begin{figure}
\centerline{\includegraphics[width=\columnwidth]{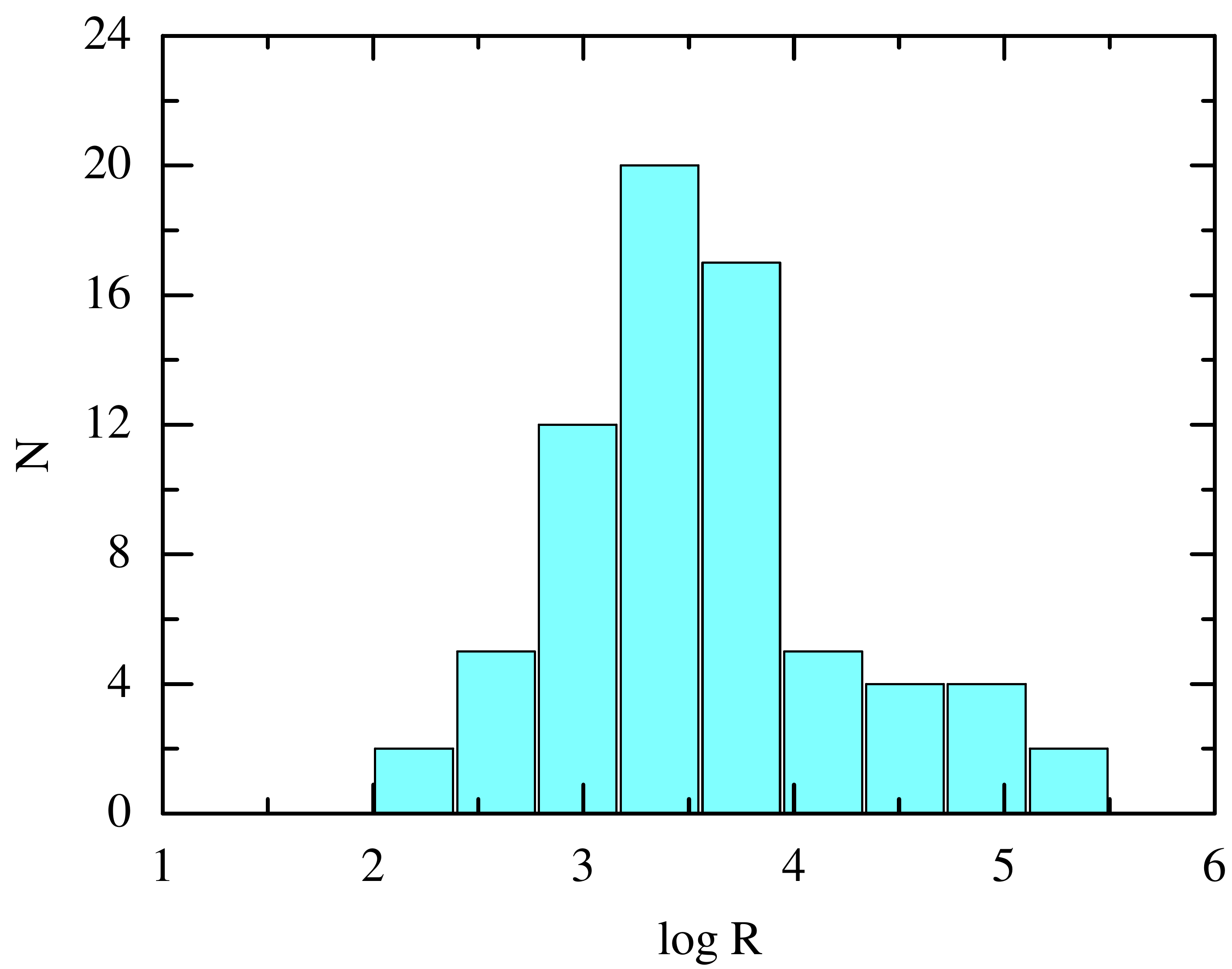}}
\caption{Radio loudness distribution.}
\label{fig11}
\end{figure}

Fig.~\ref{fig11} presents the radio loudness distribution. The $\log~R$ value spans from 2.1 to 5.43, the median is 3.5, and the majority of the sources are highly radio loud with $\log~R > 2.5$ \citep{2019MNRAS.482.2016Z}. Only 5 of 71 quasars have $\log~R < 2.5$.  

\begin{figure}
\centerline{\includegraphics[width=\columnwidth]{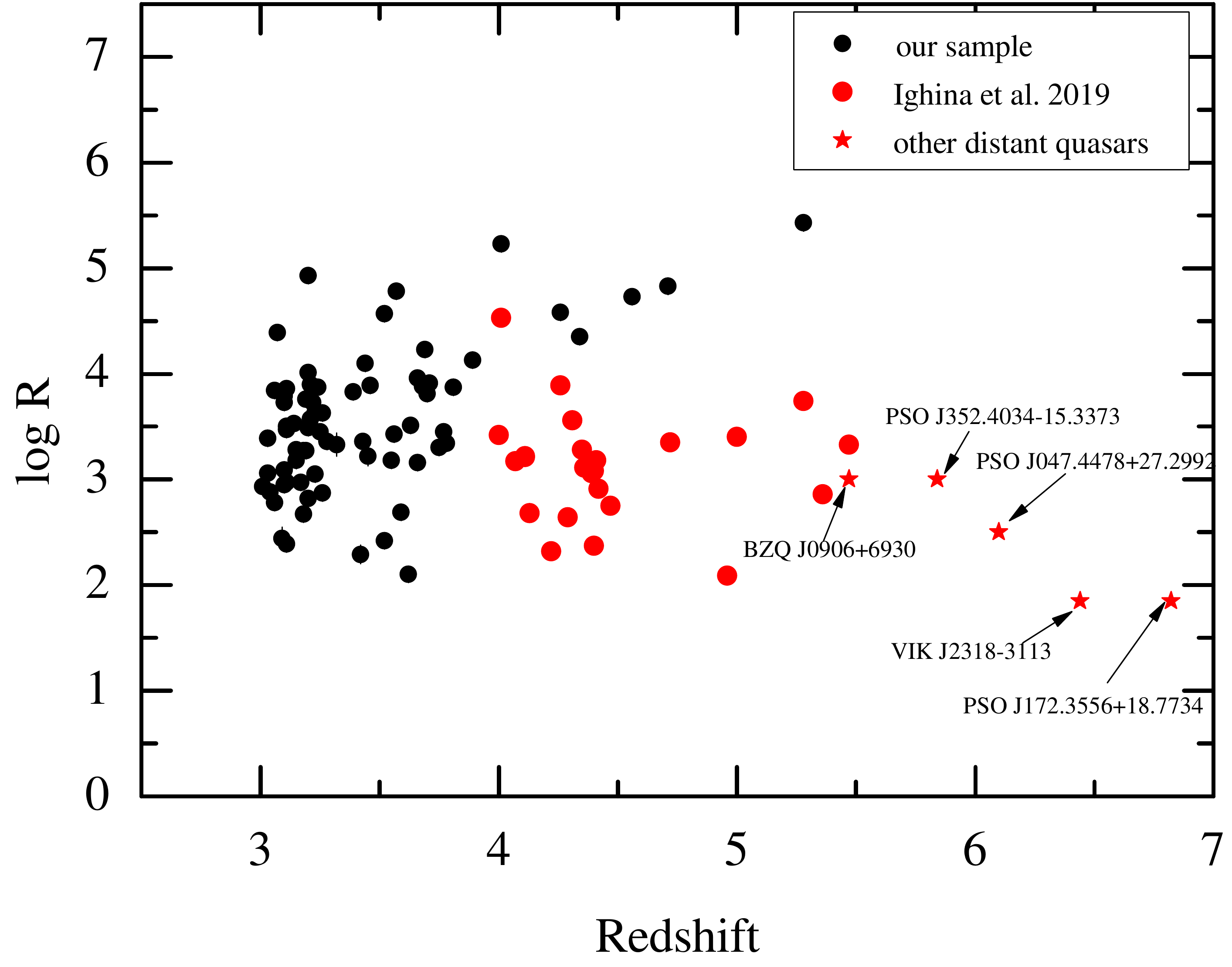}}
\caption{Redshift versus radio loudness. The sample of distant blazars from \citet{2019MNRAS.489.2732I} is shown with the red circles, and other five distant quasars from the literature are marked with the red stars.}
\label{fig12}
\end{figure}

In Fig.~\ref{fig12} we draw 24 radio-loud blazars at $z>4$ selected from the Cosmic Lens All Sky Survey (CLASS: \citealt{2019MNRAS.489.2732I}) as well as five individual sources - BZQ J0906+6930 at $z=5.47$ \citep{2004ApJ...610L...9R}, PSO J352.4034-15.3373 at $z=5.84$ \citep{2018ApJ...861L..14B}, PSO J047.4478+27.2992 at $z=6.1$ \citep{2020A&A...635L...7B}, VIK J2318-3113 at $z=6.44$ \citep{2021arXiv210111371I}, and PSO J172.3556+18.7734 at $z=6.82$ \citep{2021ApJ...909...80B}, - to compare with our results. 
The sample of \citet{2019MNRAS.489.2732I} was selected with a lower limit of 30 mJy at 1.4 GHz, therefore it contains fainter radio sources with lower radio-loudness values than our sample, which is reflected in Fig.~\ref{fig12}. We also discuss selection effects in the following section.

\section{Discussion}

We have classified 46 per cent spectra of $z\geq 3$ quasars as PS type, which is quite a large fraction of bright distant quasars. Thus, we confirm the earlier result of \citet{1990A&AS...84..549O} who revealed that seven of the 14 quasars studied at $z\geq 3$ are GPS. Also, \citet{1998PASP..110..493O} expects about 40 per cent of all bright radio sources to be PS type (GPS and CSS). 

If a PS source has a peak between 0.5 and 10 GHz, then it is classified as GPS; however, if the spectrum has a peak around 1 GHz, the spectral indices of the optically thin and thick parts are about 0.51 and $-$0.73, and the source is non-variable, then it is classified more precisely as a classical or canonical GPS \citep{1997A&A...321..105D}. In our sample, 17 sources (17 per cent) fit the criteria of classical GPS. 12 of them are already known as GPS or MPS in \citet{1997A&A...321..105D}, \citet{1998PASP..110..493O}, and \cite{2017ApJ...836..174C}. For the other five, further study of their variability (Sotnikova 2021, in preparation) will help classify them more precisely. The variability is a crucial parameter when distinguishing between PS quasars and classical GPS \citep{2020arXiv200902750O}.

Several studies \citep{2005A&A...435..839T,2012A&A...544A..25M,2013AstBu..68..262M,2019AstBu..74..348S} have shown that only 2--3 per cent of PS objects are classical GPS, whereas other GPS sources could be, in fact, the blazars that temporarily have broader peaked spectra during their flares  \citep{2011A&A...536A..14P}. 50 per cent of PS objects in the sample (24 of 47) are FSRQs .

We compared the spectra of GPS quasars in the samples from \citet{2012A&A...544A..25M,2013AstBu..68..262M} and \citet{2019AstBu..74..348S}. There is no difference between averaged optically thick and thin spectral indices within the measurement uncertainties (Table~\ref{tab:comparing}). This can mean that the high redshift GPS have, on average, the same spectral indices as the low- and medium-redshift GPS, and there is no evidence for the evolution of the radio spectrum.

Among 47 PS objects, we determined that 13 (13 per cent) MPSs in the sample that are the sources with a peak below 1 GHz in the observer’s frame. They are also interesting as young compact (5--10 mas) radio sources \citep{2015MNRAS.450.1477C,2016MNRAS.459.2455C,2017ApJ...836..174C}, possibly evolved versions of GPS sources \citep{2010MNRAS.408.2261K}. 
On the whole, MPS sources are a combination of PS sources, the peak frequency of which have been shifted to low frequencies due to cosmological redshift \citep{2021A&ARv..29....3O}. Two of our 13 MPSs had already been identified as MPSs in \citet{2017ApJ...836..174C}. We note that 15 objects of the sample with the steep spectrum could also be MPS candidates, and additional low-frequency ($<$1 GHz) data are needed to obtain their radio spectra at megahertz frequencies and define a possible peak.

\citet{1957MNRAS.117..680W}, \citet{1979A&A....80...13B}, \citet{1979A&AS...35..153T}, \citet{1980MNRAS.190..903L}, \citet{2000A&AS..143..303D}, \citet{2001MNRAS.326.1563J}, \citet{2010AstL...36....7V}, and \citet{2014A&A...569A..52S} reported a correlation between the observed spectral index and redshift and also proposed to use the presence of an ultra-steep spectral index as a method for selecting high-redshift radio sources. However, this correlation was detected mostly for radio galaxies, while more and more distant quasars with flat radio spectra have been revealed later. The steeper spectra for distant quasars could be explained by selection effects and higher inverse Compton losses due to the increased photon energy density of the cosmic microwave background \citep{2018MNRAS.480.2726M}. We have no quasars with the ultra-steep spectrum in the sample.

The number of objects in our sample decreases with increasing redshift, whereas the radio loudness increases. This means that our initial criterion of $S_{1.4}\geq100$ mJy selects only the most radio-loud quasars at high redshifts. There are other radio-loud quasars underrepresented in our study because of their low radio flux density at $\nu_{\rm obs} = 1.4$ GHz. The quasar radio-loud fraction was investigated in several studies, and it was shown that it remains about $\sim$ 10 per cent in optically selected quasar samples up to redshift $z \sim 6$ \citep{1989AJ.....98.1195K,2000AJ....119.1526S,2021ApJ...908..124L}.
When higher redshifts are considered, the lower-frequency observations are needed to detect the radio emission because of the steep spectrum at MHz--GHz frequencies. We think that the artificial trend we obtained in the $\log R$--$z$ plane will be eliminated if the selection criterion is changed toward a lower flux density limit at 1.4 GHz, or if a lower frequency survey is applied to the high-redshift sample construction.

\begin{table*}
\caption{\label{tab:comparing} Comparison of spectral indices of quasars with the classical GPS features in different samples: $\alpha_{\rm below}$ and $\alpha_{\rm above}$ are the median spectral indices below and above the turnover, respectively.}
\centering
\begin{tabular}{|c|c|c|c|c|}
\hline
   $z$  & N &  $\alpha_{\rm below}$ & $\alpha_{\rm above}$ &  Reference \\[3pt]
\hline
0.14--4.56 & 45 & 0.7 (0.1) & $-$0.7 (0.1) & {\cite{2012A&A...544A..25M}} \\
0--5       & 43 & 0.9 (0.1) & $-$0.6 (0.1) & \cite{2013AstBu..68..262M} \\
           &    &            &            & $S_{5}\geq 0.2$ Jy \\
0--5       & 71 & 0.7 (0.2) & $-$0.7 (0.2) & \cite{2019AstBu..74..348S}   \\ 
           &    &            &            & $S_{5}\geq 0.2$ Jy \\
3--5       & 12 & 0.7 (0.2) & $-$0.7 (0.3) & This study  \\ 
           &    &            &            & $S_{1.4}\geq 0.1$ Jy \\
\hline
\end{tabular}
\end{table*}

\section{Summary}

We present the results of multifrequency RATAN observations in 2017--2020 of a complete sample of bright quasars at $z \gid 3$. The flux densities were measured at frequencies of 1.1, 2.3, 4.7, 8.2, 11.2, and 22 GHz quasi-simultaneously with uncertainties of 9--31 per cent. The detection rate is 100, 89, and 46 per cent at 4.7, 11.2, and 22 GHz, respectively. The main conclusions are the following:
\begin{enumerate}
\item The sample has peaked and flat radio spectra (46 per cent and 24 per cent, respectively). There are 15 per cent quasars with a steep spectrum ($-1.1<\alpha<-0.5$), indicating the dominance of bright compact core emission and insignificant contribution of extended optically thin kpc-scale components in observed radio spectra. There are no quasars with ultra-steep radio spectra ($\alpha\leq-1.1$) in the sample. We have not found any significant correlation between the redshifts and spectral indices.
\item Eight new MPSs (J0214$+$0157, J0232$+$2317, J0525$-$3343, J0624$+$3856, J0905$+$0410, J1045$+$3142, J1418$+$4250, J2019$+$1127) and 16 new GPS (Table~\ref{tab:PS}) candidates are suggested. Further study of their variability and additional low-frequency observations are needed to classify them precisely. 
\item The radio luminosity $L_{4.7}$ has been calculated with a median value of $\sim 2 \times 10^{44}$ erg s$^{-1}$ for the sample.
\item Using new radio data, we have estimated the radio loudness for 71 objects with a mean value of 3.5. The $\log~R$ values span from 2.1 to 5.43. The majority of the quasars are highly radio-loud with $\log~R >~2.5$, and only five quasars have $\log~R <~2.5$. 
\end{enumerate}

We are preparing a follow-up work with a further study of the sample, including assessing the variability and notes on individual objects (Sotnikova, in preparation).

\section*{Acknowledgements}
We thank the referee for providing useful suggestions and comments that significantly improved the article. This work was supported in the framework of the national project ``Science'' by the Ministry of Science and Higher Education of the Russian Federation under the contract 075-15-2020-778. The observations were carried out with the RATAN-600 scientific facility. This research has made use of the CATS database, operated at SAO RAS, Russia.
The research has made use of the NASA/IPAC Extragalactic Database (NED), which is operated by the Jet Propulsion Laboratory, California Institute of Technology, under contract with the National Aeronautics and Space Administration. We used of the SIMBAD database, operated at CDS, Strasbourg, France.

\section*{Data Availability}
The data underlying this article are available in the article and in its online supplementary material. The measured flux densities are distributed in the VizieR Information System and partly published in the BL Lac database available at the Special Astrophysical Observatory website.

\bibliographystyle{mnras} %
\bibliography{manuscript} %

\appendix
\onecolumn  
\clearpage
\newpage

\begin{small}
\begin{longtable}{|l|c|c|c|c|c|c|c|c|}
\caption{\label{tab:param}The sample parameters: source name, redshift $z$, average flux density $S_{4.7}$ and radio luminosity $L_{4.7}$, spectral indices $\alpha_{\rm low}$ and $\alpha_{\rm high}$, radio-loudness $\log~R$, radio-spectrum type, and blazar type. The redshifts are taken from the NED database, and other parameters presented in Columns 3--8 are estimations based on RATAN-600 measurements.}\\
\hline
\multirow{2}{*}{NVSS name} &  \multirow{2}{*}{$z$} & $S_{4.7}$, &  $L_{4.7}\times10^{44}$,  &  \multirow{2}{*}{$\log~R$} & \multirow{2}{*}{$\alpha_{\rm low}$} & \multirow{2}{*}{$\alpha_{\rm high}$} & \multirow{2}{*}{Sp. type} & \multirow{2}{*}{Blazar type} \\
          &      &        (Jy) &                   (erg s$^{-1}$)   &    &  & &      &            \\
 1 & 2 & 3 & 4 & 5 & 6 & 7 & 8 & 9 \\
\hline
\endfirsthead
\caption[]{continued.}\\
\hline
\multirow{2}{*}{NVSS name} & \multirow{2}{*}{$z$} & $S_{4.7}$, &  $L_{4.7}\times10^{44}$,  & \multirow{2}{*}{$\log~R$} & \multirow{2}{*}{$\alpha_{\rm low}$} & \multirow{2}{*}{$\alpha_{\rm high}$} & \multirow{2}{*}{Sp. type} & \multirow{2}{*}{Blazar type} \\
          &      &   (Jy)    &   (erg s$^{-1}$)       &                    &  & &          &    \\
 1 & 2 & 3 & 4 & 5 & 6 & 7 & 8 & 9 \\
\hline
\endhead
\hline
\endfoot
000108$+$191434     &   3.10 & $0.15\pm0.01$   &  $1.38\pm0.17$   &        & $-0.87\pm0.11$ & $+0.16\pm0.01$ & upturn  & FSRQ \\ 
000657$+$141546     &   3.20 & $0.14\pm0.01$   &  $1.27\pm0.13$   &   $2.82\pm0.04$ & $+0.02\pm0.02$ & $-0.21\pm0.01$ & peaked  & \\
004858$+$064005     &   3.58 & $0.17\pm0.01$   &  $9.07\pm1.13$   &        & $+0.05\pm0.03$ & $-0.95\pm0.01$ & peaked  &  \\
010012$-$270852     &   3.52 & $0.12\pm0.01$   &  $0.74\pm0.11$   &   $2.42\pm0.05$ & $+0.41\pm0.04$ & $-0.47\pm0.01$ & peaked  &  \\
012100$-$280622     &   3.11 & $0.17\pm0.01$   &   $1.74\pm0.12$  &        & $+1.45\pm0.15$ & $-0.08\pm0.01$ & peaked & FSRQ\\ 
013113$+$435813     &   3.12 & $0.03\pm0.01$   &   $0.54\pm0.10$   &        & $-1.16\pm0.02$ & $-0.12\pm0.08$ & complex  & \\ 
014844$+$421519     &   3.24 & $0.11\pm0.01$   &   $1.68\pm0.20$  &        & $-0.27\pm0.02$ & $+0.20\pm0.01$ & upturn & FSRQ \\
015106$+$251729     &   3.10 & $0.13\pm0.01$   &   $1.41\pm0.12$  &  $3.73\pm0.03$  & $-0.05\pm0.01$ & $-0.05\pm0.01$ & flat & \\ 
020346$+$113445     &   3.63 & $0.65\pm0.03$   &   $7.30\pm0.69$  &  $3.51\pm0.04$  & $+0.29\pm0.01$ & $-0.51\pm0.01$ & peaked$^*$ & FSRQ \\
021435$+$015703     &   3.28 & $0.08\pm0.01$   &   $1.71\pm0.22$  &  $3.36\pm0.06$  & $+0.34\pm0.09$ & $-0.45\pm0.01$ & peaked   & \\
023220$+$231757     &   3.42 & $0.31\pm0.02$   &   $5.86\pm0.64$  &        & $+0.87\pm0.08$ & $-0.37\pm0.01$ & peaked$^*$ & \\
024611$+$182330     &   3.59 & $0.17\pm0.01$   &   $3.32\pm0.24$  &        & $+0.49\pm0.02$ & $-0.35\pm0.01$ & peaked & FSRQ \\ 
025759$+$433838     &   4.06 & $0.31\pm0.02$   &   $2.59\pm0.25$  &        & $+0.36\pm0.01$ & $+0.04\pm0.01$ & inverted & \\
032444$-$291821     &   4.63 & $0.15\pm0.01$   &   $1.58\pm0.35$  &        & $+0.15\pm0.01$ & $+0.15\pm0.01$ & inverted &  FSRQ\\
033755$-$120404     &   3.44 & $0.29\pm0.02$   &   $4.79\pm0.42$  & $4.10\pm0.04$   & $-0.32\pm0.01$ & $-0.27\pm0.01$ & flat & FSRQ\\
033900$-$013318     &   3.19 & $0.37\pm0.02$   &   $5.43\pm0.49$  & $3.76\pm0.04$   & $-0.03\pm0.01$  & $-0.40\pm0.01$ & flat & FSRQ\\
035424$+$044107     &   3.26 & $0.39\pm0.02$   &   $3.97\pm0.38$  &        & $+0.04\pm0.02$  & $-0.36\pm0.01$ & complex  &  \\
042457$+$080517     &   3.09 & $0.19\pm0.01$   &   $1.59\pm0.16$  &        & $+0.11\pm0.02$  & $+0.19\pm0.01$ & inverted & FSRQ \\
042835$+$173223     &   3.32 & $0.15\pm0.01$   &   $1.98\pm0.13$  &        & $-0.09\pm0.01$  & $-0.09\pm0.01$ & flat & FSRQ \\
052506$-$233810     &   3.1  & $0.74\pm0.03$   &   $3.85\pm0.38$  & $3.09\pm0.04$   & $+0.60\pm0.01$  & $+0.34\pm0.01$ & inverted & bl.un. \\ 
052506$-$334305     &   4.41 & $0.40\pm0.01$   &   $33.80\pm4.45$ &        & $+0.43\pm0.02$  & $-0.96\pm0.02$ & peaked & FSRQ\\ 
053954$-$283955     &   3.10 & $0.99\pm0.10$    &   $5.65\pm0.56$ & $3.79\pm0.04$   & $+0.32\pm0.01$  & $-0.32\pm0.01$ & peaked$^{*}$ & FSRQ  \\ 
062419$+$385648     &   3.46 & $0.61\pm0.03$   &   $11.80\pm1.18$ & $3.89\pm0.04$   & $+0.40\pm0.01$  & $-0.33\pm0.01$ & peaked & FSRQ\\ 
064632$+$445116     &   3.39 & $2.27\pm0.10$    &   $10.80\pm1.00$& $3.83\pm0.04$   & $+0.52\pm0.01$  & $-0.62\pm0.01$ & peaked & FSRQ \\ 
073357$+$045614     &   3.01 & $0.53\pm0.03$   &   $3.66\pm0.37$  &        & $+0.39\pm0.01$  & $-0.31\pm0.01$ & peaked & FSRQ \\ 
075141$+$271632     &   3.20 & $0.20\pm0.01$   &   $5.29\pm0.62$  & $4.93\pm0.05$   & $+1.02\pm0.03$  & $-0.75\pm0.01$ & peaked &  \\ 
075303$+$423131     &   3.59 & $0.41\pm0.02$   &   $1.18\pm0.17$  & $2.69\pm0.05$   & $+0.87\pm0.04$  & $-0.61\pm0.01$ & peaked & \\ 
083322$+$095941     &   3.75 & $0.09\pm0.01$   &   $2.12\pm0.23$  & $3.30\pm0.05$   & $-0.33\pm0.01$  & $-0.33\pm0.01$ & flat &  \\
083910$+$200207     &   3.03 & $0.12\pm0.01$   &   $0.64\pm0.07$  & $3.06\pm0.04$   & $+0.36\pm0.01$  & $-0.25\pm0.01$ & peaked &  \\ 
084715$+$383110     &   3.18 & $0.11\pm0.01$   &   $1.08\pm0.10$  & $2.67\pm0.04$   & $+0.07\pm0.01$ & $+0.07\pm0.01$ & inverted & \\
090549$+$041010     &   3.15   & $0.09\pm0.01$ &   $1.98\pm0.23$  & $3.28\pm0.05$   & $+0.57\pm0.06$  & $-0.53\pm0.01$ & peaked & \\ 
090915$+$035443     &   3.20   & $0.13\pm0.01$ &   $1.61\pm0.14$  & $3.49\pm0.04$   & $-0.35\pm0.01$   & $-0.10\pm0.01$ &  flat & \\
091551$+$000712     &   3.07   & $0.21\pm0.01$ &   $3.44\pm0.35$  & $4.39\pm0.05$   & $-0.28\pm0.01$   & $-0.28\pm0.01$ &  flat & FSRQ\\
093337$+$284532     &   3.42   & $0.05\pm0.01$ &   $0.47\pm0.01$  & $2.29\pm0.09$   & $+0.17\pm0.02$   & $-0.67\pm0.01$ &  peaked &  \\ 
094113$+$114532     &   3.19   & $0.12\pm0.01$ &   $1.94\pm0.23$  & $3.27\pm0.05$   & $-0.51\pm0.01$   & $-0.51\pm0.01$ &  flat & \\
101644$+$203747     &   3.11   & $0.55\pm0.03$ &   $8.93\pm0.79$  & $3.86\pm0.04$ & $-0.07\pm0.01$ & $-0.52\pm0.01$ &  complex & FSRQ \\
102010$+$104003     &   3.15   & $0.15\pm0.01$ &   $4.77\pm0.40$  & $3.18\pm0.04$ & $-0.63\pm0.01$   & $-0.94\pm0.01$ &  steep &   \\
102107$+$220922     &   4.26   & $0.12\pm0.01$ & $2.17\pm0.23$    & $4.58\pm0.05$ & $-0.51\pm0.01$   & $-0.17\pm0.01$ &  flat & \\
102623$+$254259     &   5.28   & $0.11\pm0.01$ &   $6.31\pm0.60$  & $5.43\pm0.04$ & $+0.11\pm0.03$   & $-0.50\pm0.01$ &  peaked & FSRQ \\
102645$+$365826     &   3.25   & $0.19\pm0.01$ &   $2.25\pm0.19$  & $3.45\pm0.03$ & $-0.08\pm0.01$   & $-0.27\pm0.01$ &  flat  &  \\
102838$-$084438     &   4.27   & $0.10\pm0.01$ &   $2.25\pm0.21$  &       & $-0.11\pm0.01$   & $-0.11\pm0.01$ &  flat & FSRQ \\
103626$+$132652     &   3.09   & $0.04\pm0.01$ &   $0.94\pm0.22$  &  $2.44\pm0.11$ & $-0.89\pm0.01$   & $-0.53\pm0.01$ &  steep  &  \\
104523$+$314232     &   3.23   & $0.08\pm0.01$ &   $1.72\pm0.25$  &  $3.05\pm0.06$ & $+0.25\pm0.03$   & $-0.56\pm0.01$ & peaked  &  \\ 
110147$+$001039     &   3.69   & $0.06\pm0.01$ &   $2.84\pm0.46$  &  $4.23\pm0.07$ & $-0.74\pm0.01$   & $-0.78\pm0.01$ & steep & \\
112500$+$333858     &   3.43   & $0.08\pm0.01$ &   $2.01\pm0.27$  &  $3.36\pm0.06$ & $-0.78\pm0.01$   & $-0.55\pm0.01$ & steep &  \\
112851$+$232617     &   3.04   & $0.10\pm0.01$ &   $1.63\pm0.18$  &  $2.88\pm0.05$ & $-0.46\pm0.01$   & $-0.46\pm0.01$ & flat  & \\
115016$+$433205     &   3.03   & $0.11\pm0.01$ &   $1.42\pm0.10$  &  $3.39\pm0.03$ & $+0.01\pm0.01$   & $+0.01\pm0.01$ & inverted & FSRQ \\
123055$-$113909     &   3.52   & $0.18\pm0.01$ &   $5.67\pm0.59$  &  $4.57\pm0.05$ & $+0.34\pm0.04$   & $-0.56\pm0.01$ & peaked & FSRQ\\ 
124209$+$372006     &   3.81   & $0.56\pm0.02$ &   $4.23\pm0.43$  &  $3.87\pm0.04$ & $+0.26\pm0.01$   & $-0.30\pm0.01$ & peaked$^*$ &  \\
130122$+$190353     &   3.10   & $0.08\pm0.01$ &   $1.63\pm0.26$  &       & $+0.13\pm0.01$   & $-0.68\pm0.01$ & peaked & \\ 
134022$+$375443     &   3.11   & $0.30\pm0.02$ &   $0.54\pm0.05$  &  $2.39\pm0.04$ & $+1.07\pm0.01$ & $-1.02\pm0.01$ & peaked & \\ 
135406$-$020603     &   3.70   & $0.70\pm0.04$ &   $4.80\pm0.49$  &  $3.81\pm0.04$ & $+0.46\pm0.01$ & $-0.35\pm0.01$ & peaked & FSRQ\\ 
135646$-$110130     &   3.01   & $0.23\pm0.01$ &   $1.53\pm0.18$  &  $2.93\pm0.05$ & $+0.13\pm0.01$ & $+0.13\pm0.01$ & inverted & FSRQ\\
135652$+$291817     &   3.24   & $0.05\pm0.01$ &   $1.63\pm0.25$  &  $3.87\pm0.07$ & $-0.57\pm0.01$ & $-0.57\pm0.01$ & steep  & \\
135706$-$174402     &   3.14   & $0.65\pm0.04$ &   $7.20\pm0.75$  &  $3.53\pm0.04$ & $+0.28\pm0.01$ & $-0.42\pm0.01$ & peaked$^{*}$  & FSRQ \\ 
140135$+$151326     &   3.23   & $0.04\pm0.01$ &   $0.38\pm0.10$  &  $3.63\pm0.12$ & $+0.05\pm0.01$ & $-0.78\pm0.02$ & peaked &  \\ 
140501$+$041535     &   3.20   & $0.73\pm0.03$ &   $9.65\pm0.89$  &  $4.01\pm0.04$ & $+0.01\pm0.01$ & $-0.43\pm0.01$ & peaked$^*$  & FSRQ \\
141152$+$430024     &   3.21   & $0.09\pm0.01$ &   $1.30\pm0.14$  &  $3.57\pm0.05$ & $-0.21\pm0.01$ & $-0.21\pm0.01$ & flat  & \\
141300$+$394745     &   3.71   & $0.05\pm0.01$ &   $1.44\pm0.27$  &  $3.91\pm0.08$ & $-0.55\pm0.01$ & $-0.44\pm0.01$ & flat  &  \\
141318$+$450522     &   3.11   & $0.16\pm0.01$ &   $1.40\pm0.10$  &  $3.47\pm0.03$ & $+0.04\pm0.01$ & $+0.04\pm0.01$ & inverted  & FSRQ \\
141821$+$425020     &   3.45   & $0.05\pm0.01$ &   $0.38\pm0.08$  &  $3.22\pm0.09$ & $+0.32\pm0.01$ & $-0.77\pm0.01$ & peaked &  \\ 
142107$-$064356     &   3.68   & $0.22\pm0.01$ &   $5.33\pm0.52$  &  $3.88\pm0.04$ & $-0.65\pm0.01$ & $-0.38\pm0.01$ & complex  & FSRQ\\
142438$+$225600     &   3.62   & $0.61\pm0.03$ &   $2.52\pm0.28$  &  $2.10\pm0.05$ & $+0.62\pm0.01$ & $-0.59\pm0.01$ & peaked  & FSRQ\\ 
143023$+$420436     &   4.71   & $0.12\pm0.01$ &   $3.88\pm0.27$  &  $4.83\pm0.03$ & $-0.13\pm0.01$ & $-0.13\pm0.01$ & flat & FSRQ \\
144516$+$095836     &   3.55   & $0.98\pm0.04$ &   $4.12\pm0.43$  &  $3.18\pm0.04$ & $+0.68\pm0.01$ & $-0.87\pm0.01$ & peaked & FSRQ\\
145722$+$051922     &   3.17   & $0.07\pm0.01$ &   $1.16\pm0.16$  &  $2.97\pm0.06$ & $-0.30\pm0.01$ & $-0.30\pm0.01$ & flat   &  \\
145805$+$085529     &   3.06   & $0.06\pm0.01$ &   $2.10\pm0.35$  &  $3.84\pm0.07$ & $-0.53\pm0.01$ & $-0.87\pm0.01$ & steep  &  \\
145927$+$325359     &   3.32   & $0.05\pm0.01$ &   $1.54\pm0.39$  &  $3.33\pm0.11$ & $-0.67\pm0.01$ & $-0.85\pm0.01$ & steep  &  \\
150328$+$041949     &   3.66   & $0.19\pm0.01$ &   $2.23\pm0.15$  &  $3.16\pm0.03$ & $+0.21\pm0.01$ & $-0.29\pm0.01$ & peaked  &  \\
152117$+$175601     &   3.06   & $0.13\pm0.01$ &   $0.81\pm0.08$  &  $2.78\pm0.04$ & $+0.40\pm0.03$ & $-0.87\pm0.01$ & peaked  & \\ 
152219$+$211957     &   3.22   & $0.08\pm0.01$ &   $3.35\pm0.48$  &  $3.73\pm0.06$ & $-0.99\pm0.01$ & $-1.04\pm0.01$ & steep     & \\
153815$+$001905     &   3.49   & $0.32\pm0.02$ &   $24.00\pm2.50$  &       & $+0.38\pm0.01$ & $-0.77\pm0.01$ & peaked &    \\ 
155930$+$030447     &   3.89   & $0.43\pm0.02$ &   $4.97\pm0.50$  &  $4.13\pm0.04$ & $+0.11\pm0.01$ & $-0.43\pm0.01$ & peaked & FSRQ\\ 
160002$+$041256     &   3.11   & $0.16\pm0.01$ &   $2.07\pm0.14$  &  $3.50\pm0.03$ & $-0.19\pm0.01$ & $-0.19\pm0.01$ & flat  & \\
160608$+$312504     &   4.56   & $0.64\pm0.03$ &   $1.73\pm0.22$  &  $4.73\pm0.05$ & $+1.10\pm0.02$ & $-0.57\pm0.01$ & peaked &  \\ 
161005$+$181143     &   3.11   & $0.07\pm0.01$ &   $2.20\pm0.35$  &  $2.97\pm0.06$ & $-0.25\pm0.01$ & $-0.79\pm0.02$ & steep & \\
161637$+$045932     &   3.21   & $1.03\pm0.04$ &   $8.63\pm0.80$  &  $3.90\pm0.04$ & $+0.64\pm0.01$ & $-0.56\pm0.01$ & peaked$^*$  & FSRQ \\
163257$-$003321     &   3.42   & $0.15\pm0.01$ &   $2.17\pm0.16$  &       & $-0.19\pm0.01$ & $-0.19\pm0.01$ & flat  & FSRQ \\ 
165519$+$324241     &   3.18   & $0.06\pm0.01$ &   $2.06\pm0.33$  &  $3.27\pm0.07$ & $-0.80\pm0.01$ & $-0.80\pm0.01$ & steep  & \\
165543$+$194847     &   3.26   & $0.11\pm0.01$ &   $1.97\pm0.20$  &  $3.63\pm0.04$ & $-0.10\pm0.01$ & $-0.36\pm0.01$ & flat  &  \\
165844$-$073917     &   3.74   & $0.83\pm0.03$ &   $19.0\pm1.77$  &       & $+0.16\pm0.01$ & $-0.86\pm0.01$ & peaked$^*$ & FSRQ \\
171521$+$214532     &   4.01   & $0.16\pm0.01$ &   $6.26\pm0.44$  &  $5.23\pm0.03$ & $-0.66\pm0.01$ & $-0.66\pm0.01$ & steep & \\
174020$+$350048     &   3.22   & $0.06\pm0.01$ &   $1.64\pm0.29$  &       & $-0.69\pm0.01$ & $-0.69\pm0.01$ & steep  & \\
184057$+$390046     &   3.09   & $0.18\pm0.01$ &   $3.58\pm0.33$  &       & $+0.27\pm0.01$ & $-0.32\pm0.01$ & peaked & FSRQ \\ 
193957$-$100241     &   3.78   & $0.59\pm0.02$ &   $12.00\pm1.08$  & $3.34\pm0.04$  & $-0.17\pm0.01$ & $-0.73\pm0.01$ & complex   & FSRQ \\
200324$-$325144     &   3.77   & $0.68\pm0.03$ &   $11.70\pm1.07$ & $3.45\pm0.04$  & $+0.40\pm0.01$ & $-0.74\pm0.01$ & peaked  & FSRQ\\ 
201918$+$112712     &   3.27   & $0.08\pm0.01$ &   $3.14\pm0.40$  &       & $+0.71\pm0.01$ & $-0.84\pm0.01$ & peaked  &    \\ 
204124$+$185502     &   3.05   & $0.17\pm0.01$ &   $5.15\pm0.62$  &       & $-0.83\pm0.01$ & $-0.83\pm0.01$ & steep   &  \\
204257$-$222326     &   3.63   & $0.09\pm0.01$ &   $6.69\pm0.86$  &       & $+0.13\pm0.01$ & $-0.60\pm0.02$ & peaked  & FSRQ \\
204310$+$125513     &   3.27   & $0.14\pm0.02$ &   $2.44\pm0.30$  &       & $-0.18\pm0.01$ & $-0.18\pm0.01$ & flat  & FSRQ\\
205051$+$312727     &   3.18   & $0.57\pm0.03$ &   $9.89\pm0.89$  &       & $+0.15\pm0.01$ & $-0.39\pm0.01$ & peaked  & FSRQ\\ 
212912$-$153840     &   3.26   & $1.46\pm0.10$ &   $6.07\pm0.50$  & $2.87\pm0.04$  & $+0.56\pm0.01$ & $-0.61\pm0.01$ & peaked$^{*}$ & FSRQ\\ 
213412$-$041910     &   4.34   & $0.22\pm0.01$ &   $5.15\pm0.58$  & $4.35\pm0.05$  & $-0.31\pm0.01$ & $-0.31\pm0.01$ & flat & \\
221748$+$022010     &   3.57   & $0.32\pm0.02$ &   $10.8\pm1.17$  & $4.78\pm0.05$  & $-0.65\pm0.01$ & $-0.52\pm0.01$ & steep & FSRQ\\
221935$-$271903     &   3.63   & $0.21\pm0.01$ &   $3.08\pm0.32$  &       & $-0.14\pm0.01$ & $-0.14\pm0.01$ & flat  & FSRQ \\
222536$+$204015     &   3.56   & $0.13\pm0.01$ &   $3.79\pm0.40$  & $3.43\pm0.05$  & $-0.72\pm0.01$ & $-0.72\pm0.01$ & steep  & \\
224800$-$054118     &   3.29   & $0.20\pm0.01$ &   $5.10\pm0.53$  &       & $+0.02\pm0.01$ & $-0.57\pm0.01$ & peaked & bl.un. \\
225153$+$221737     &   3.66   & $0.11\pm0.01$ &   $4.75\pm0.47$  & $3.96\pm0.04$  & $-0.36\pm0.01$ & $-0.36\pm0.01$ & flat   & FSRQ\\
231448$+$020151     &   4.11   & $0.07\pm0.01$ &   $2.65\pm0.40$  &       & $+0.06\pm0.01$ & $-0.53\pm0.01$ & peaked  & \\
231643$-$334912     &   3.1    & $0.48\pm0.03$ &   $1.68\pm0.17$  & $2.95\pm0.04$  & $+0.75\pm0.02$ & $-1.58\pm0.01$ & peaked  & FSRQ \\ 
232118$-$082721     &   3.16   & $0.18\pm0.01$ &   $2.27\pm0.29$  &       & $-0.46\pm0.01$ & $-0.17\pm0.01$ & flat & FSRQ \\
234451$+$343349     &   3.05   & $0.09\pm0.01$ &   $1.62\pm0.20$  &       & $-0.50\pm0.01$ & $-0.50\pm0.01$ & flat  &  \\
\hline
\multicolumn{7}{l}{\footnotesize$^*$ sources with two components in the radio spectrum}\\
\end{longtable}
\end{small}

\begin{table*}
\caption{\label{tab:PS} Peaked spectrum (PS) sources with their peak frequencies calculated in the observer's frame (Col.3) and rest-frame (Col.4), flux density at the peak frequency (Col.5), PS type based solely on peak frequency value $\nu_{\rm peak,obs.}$ (Col.6), and the reference for previous classification of an object (Col.7).}
\centering
\begin{tabular}{|c|c|c|c|c|c|c|}
\hline
\multirow{2}{*}{NVSS name}	&	\multirow{2}{*}{$z$}	&	$\nu_{\rm peak,obs}$	&	$\nu_{\rm peak,rest}$ &	$S_{\rm peak}$	& \multirow{2}{*}{PS type} & \multirow{2}{*}{Reference} \\
 & & (GHz) & (GHz) & (Jy) & & \\
1 & 2 & 3 & 4 & 5 & 6 & 7 \\
\hline
000657$+$141546	&	3.20	    &	1.5	&	6.3 	&	0.18	&	GPS &	\\
004858$+$064005	&	3.58	&	3.9	&	17.6	&	0.22	&	GPS	& {\citet{2013AstBu..68..262M}} \\ 
010012$-$270852	&	3.52	&	1.4	&	6.3	&	0.22	&	GPS &	\\
012100$-$280622	&	3.11	&	0.3	&	1.2	&	0.20	&	MPS	& {\citet{2013AstBu..68..262M}} \\ 
020346$+$113445	&	3.63	&	3.7	&	17.1	&	1.01	&	GPS	&  {\citet{1991ApJ...380...66O}} \\ 
021435$+$015703	&	3.28	&	0.5	&	1.9	&	0.25	&	MPS	& \\
023220$+$231757	&	3.42	&	0.5	&	2.2	&	0.69	&	MPS	& \\
024611$+$182330	&	3.59	&	1.8	    &	8.3	&	0.23	&	GPS	& \\
052506$-$334305	&	4.41	&	0.9	&	4.9 	&	0.20	&	MPS	& \\
053954$-$283955	&	3.10	&	7.0	    &	28.7	&	1.08	&	GPS	& \\
062419$+$385648	&	3.46	&	0.3	    &	1.3	&	1.56	&	MPS	& \\
064632$+$445116	&	3.39	&	17.3	&	76.0	&	3.14	&	GPS &  {\citet{1991ApJ...380...66O}}\\ 
073357$+$045614	&	3.10	    &	5.8	    &	23.3	&	0.46	&	GPS	& \\ 
075141$+$271632	&	3.20  	&	0.2	&	0.8	&	1.85	&	MPS	&  {\citet{2017ApJ...836..174C}} \\
075303$+$423131	&	3.59	&	0.9	&	4.1	&	1.01	&	MPS	& {\citet{2013AstBu..68..262M}}\\ 
083910$+$200207	&	3.30	    &	1.8	    &	7.7 	&	0.14	&	GPS	& \\
090549$+$041010	&	3.15	&	0.5 	&	2.1 	&	0.32	&	MPS	& \\
093337$+$284532	&	3.42	&	1.7	    &	7.5 	&	0.11	&	GPS	& \\
102623$+$254259	&	5.28	&	0.3 	&	1.9 	&	0.46	&	MPS	& \\
104523$+$314232	&	3.23	&	0.9 	&	3.8 	&	0.20	&	MPS	& \\
123055$-$113909	&	3.52	&	1.1 	&	5.0 	&	0.44	&	GPS	&  {\citet{2013AstBu..68..262M}}\\
124209$+$372006	&	3.81	&	1.7	&	8.2 	&	0.62	&	GPS	& \\
130122$+$190353	&	3.10	    &	1.7	&	7.0 	&	0.13	&	GPS	& \\
134022$+$375443	&	3.11	&	2.7	&	11.1	&	0.46	&	GPS	& {\citet{2013AstBu..68..262M}} \\ 
135406$-$020603	&	3.70	&	2.6	&	12.2	&	0.91	&	GPS	& {\citet{1991ApJ...380...66O}} \\ 
135706$-$174402	&	3.14	&	1.8	&	7.5	    &	1.47	&	GPS	& {\citet{1991ApJ...380...66O}} \\ 
140135$+$151326	&	3.23	&	1.3	&	5.5 	&	0.11	&	GPS	& \\
140501$+$041535	&	3.20	&	8.3	&	34.9	&	0.82	&	GPS	& \\
141821$+$425020	&	3.45	&	0.9	&	4.0 	&	0.18	&	MPS	& \\
142438$+$225600	&	3.62	&	3.3	&	15.2	&	0.64	&	GPS	& {\citet{1991ApJ...380...66O}} \\ 
144516$+$095836	&	3.55	&	1.3 &	5.9 	&	3.16	&	GPS &  {\citet{1991ApJ...380...66O}} \\ 
150328$+$041949	&	3.66	&	5.6	&	26.1	&	0.17    &	GPS	& \\
152117$+$175601	&	3.60	    &	3.5	&	14.2	&	0.20	&	GPS	& \\
153815$+$001905	&	3.49	&	2.9	&	13.0	&	0.73	&	GPS	& {\citet{2013AstBu..68..262M}}\\
155930$+$030447	&	3.89	&	4.5	&	22.0	&	0.47	&	GPS	& {\citet{2017ApJ...836..174C}}\\
160608$+$312504	&	4.56	&	2.5	&	13.9	&	0.92	&	GPS	& \cite{1991ApJ...380...66O} \\
161637$+$045932	&	3.21	&	4.4	&	18.5	&	0.97	&	GPS & \citet{1991ApJ...380...66O} \\ 
165844$-$073917	&	3.74	&	5.9	&	28.0	&	0.99	&	GPS	& {\citet{1991ApJ...380...66O}} \\
184057$+$390046	&	3.90	    &	4.5	&	18.4	&	0.24	&	GPS	& {\cite{1999A&AS..135..273M}} \\
200324$-$325144	&	3.77	&	5.6	&	26.7	&	0.86	&	GPS	& {\citet{1991ApJ...380...66O}} \\ 
201918$+$112712	&	3.27	&	0.5 &	2.1 	&	0.27	&	MPS	& \\
204257$-$222326	&	3.63	&	3.5	&	16.2	&	0.19	&	GPS	& \\
205051$+$312727	&	3.18	&	1.9	&	7.9 	&	0.83	&	GPS	& \\
212912$-$153840	&	3.26	&	6.8	&	28.9	&	1.44	&	GPS	& {\citet{1991ApJ...380...66O}} \\ 
224800$-$054118	&	3.29	&	0.4	&	1.7 	&	0.91	&	MPS	& {\citet{2017ApJ...836..174C}}\\
231448$+$020151	&	4.11	&	1.7	&	8.7 	&	0.12	&	GPS	& \\
231643$-$334912	&	3.10    &	3.9	&	16.0	&	0.73	&	GPS	&  {\citet{2013AstBu..68..262M}} \\ 
\hline
\end{tabular}
\end{table*} 

\clearpage
\newpage

\begin{figure}
\centerline{\includegraphics[width=174mm]{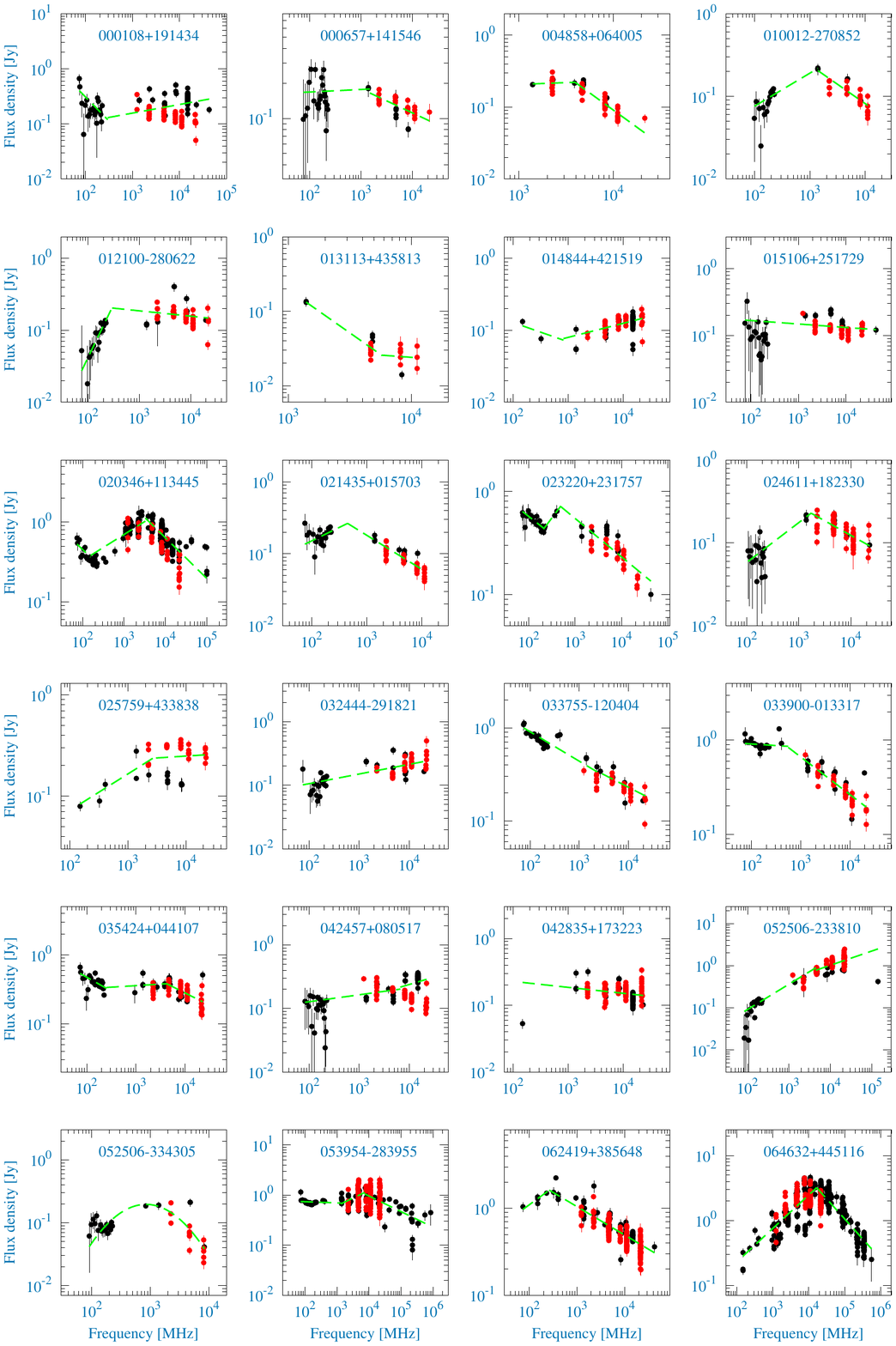}}
\caption{Broad-band radio spectra of the quasars. RATAN-600 data are shown with red circles and literature data are shown with black circles. The green dashed line represents a linear or parabolic fit to the radio spectrum.}
\label{fig:spectra}
\end{figure}

\clearpage
\newpage

\begin{figure}
\centerline{\includegraphics[width=174mm]{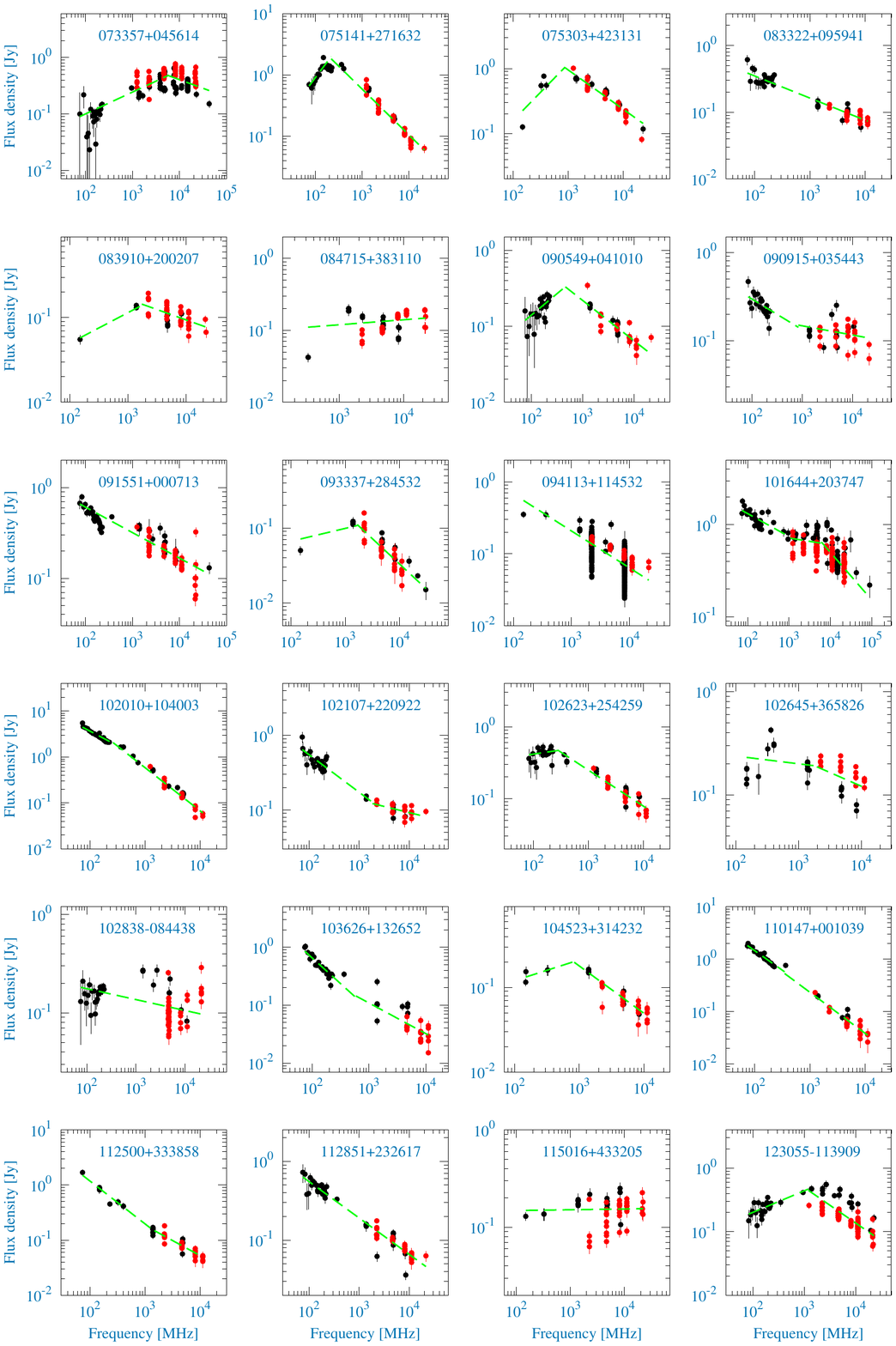}}
\caption{Broad-band radio spectra of the quasars. The description for symbols and lines is the same as in Fig.~\ref{fig:spectra}.}
\label{fig:spectra1}
\end{figure}

\clearpage
\newpage

\begin{figure}
\centerline{\includegraphics[width=174mm]{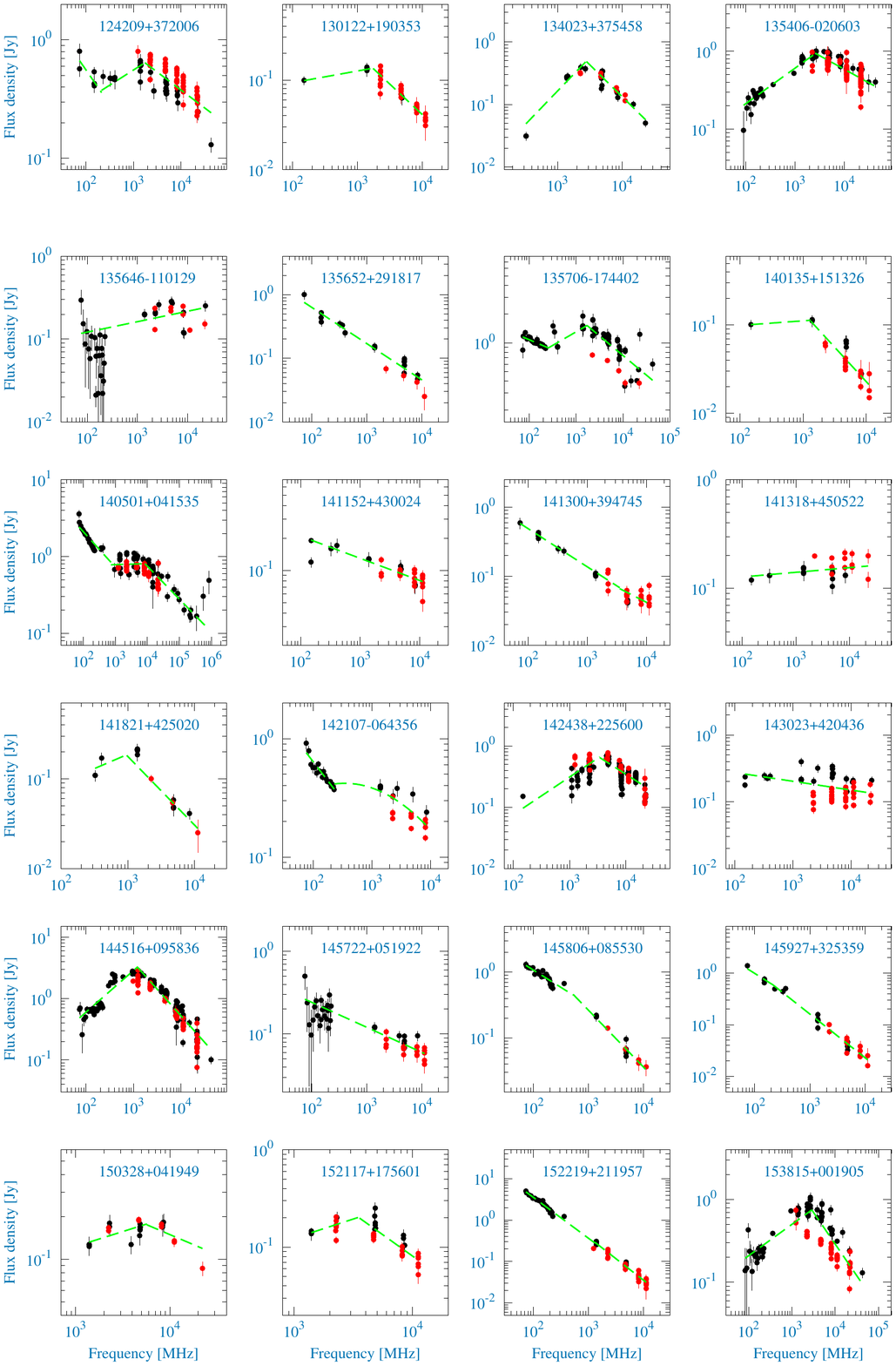}}
\caption{Broad-band radio spectra of the quasars. The description for symbols and lines is the same as in Fig.~\ref{fig:spectra}.}
\label{fig:spectra2}
\end{figure}

\clearpage
\newpage

\begin{figure}
\centerline{\includegraphics[width=174mm]{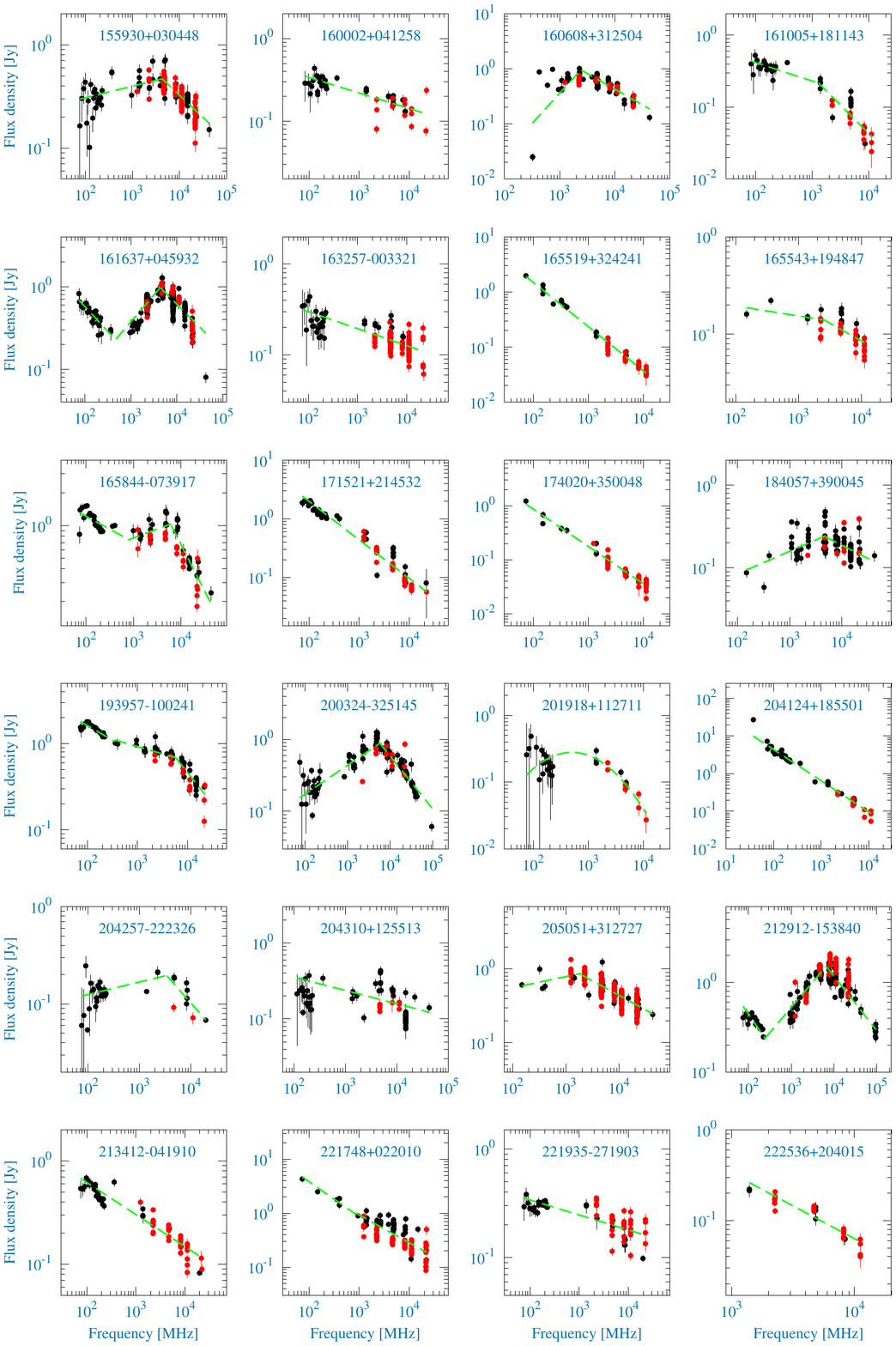}}
\caption{Broad-band radio spectra of the quasars. The description for symbols and lines are same as those in Fig.~\ref{fig:spectra}.}
\label{fig:spectra3}
\end{figure}

\clearpage
\newpage

\begin{figure}
\centerline{\includegraphics[width=142mm]{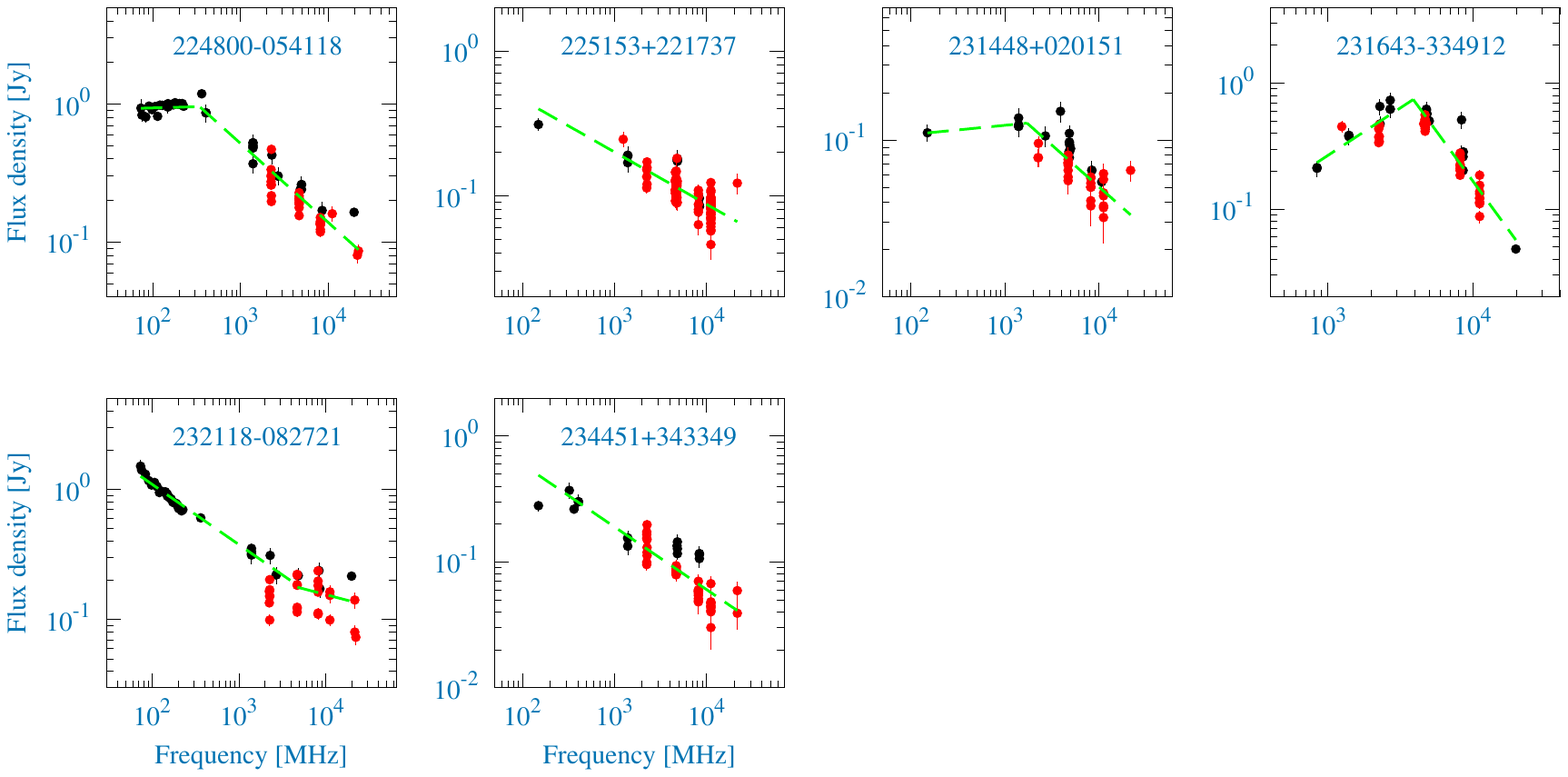}}
\caption{Broad-band radio spectra of the quasars. The description for symbols and lines is the same as in  Fig.~\ref{fig:spectra}.}
\label{fig:spectra4}
\end{figure}

\bsp	
\label{lastpage}
\end{document}